\documentclass{aa}

\usepackage{txfonts}

\usepackage{natbib,twoopt}
\usepackage[breaklinks=true]{hyperref} 
\bibpunct{(}{)}{;}{a}{}{,}             
\makeatletter
 \newcommandtwoopt{\citeads}[3][][]{\href{http://adsabs.harvard.edu/abs/#3}%
   {\def\hyper@linkstart##1##2{}%
    \let\hyper@linkend\@empty\citealp[#1][#2]{#3}}}
 \newcommandtwoopt{\citepads}[3][][]{\href{http://adsabs.harvard.edu/abs/#3}%
   {\def\hyper@linkstart##1##2{}%
    \let\hyper@linkend\@empty\citep[#1][#2]{#3}}}
 \newcommandtwoopt{\citetads}[3][][]{\href{http://adsabs.harvard.edu/abs/#3}%
   {\def\hyper@linkstart##1##2{}%
    \let\hyper@linkend\@empty\citet[#1][#2]{#3}}}
 \newcommandtwoopt{\citeyearads}[3][][]%
   {\href{http://adsabs.harvard.edu/abs/#3}
   {\def\hyper@linkstart##1##2{}%
    \let\hyper@linkend\@empty\citeyear[#1][#2]{#3}}}
\makeatother

\usepackage{graphicx}   
\usepackage{subcaption}

\usepackage{multirow, makecell}
\usepackage{booktabs}
\usepackage{xcolor}

\usepackage[normalem]{ulem} 
\DeclareUnicodeCharacter{2212}{-}

\usepackage{placeins}

\begin{document}

\title{The role of filaments in the large-scale conformity signal}

   \author{Daniela Palma\inst{1}
          \and
          Ivan Lacerna\inst{2}
          \and
          Noelia Pérez\inst{3}
          \and
          Luis Pereyra\inst{4}
          \and
          Antonio D. Montero-Dorta\inst{5}
          \and
          Constanza A. Soto-Suárez\inst{6,7}
          \and
          Laerte Sodré Jr.\inst{1}
          \and
          M. Celeste Artale\inst{8}
          \and
          Amrutha Belwadi\inst{2}
          \and
          Facundo Rodriguez\inst{9,4}
          \and
          Nelvy Choque-Challapa\inst{5}          
          }

   \institute{Instituto de Astronomia, Geofísica e Ciências Atmosféricas, Universidade de São Paulo, Rua do Matão, 1226, São Paulo, SP, 05508-090 Brazil\\
              \email{danferpalma04@gmail.com}
         \and 
         Instituto de Astronom\'ia y Ciencias Planetarias, Universidad de Atacama, Copayapu 485, Copiap\'o, Chile.
         \and
         Facultad de Ciencias Exactas, Físicas y Naturales, Departamento de Geofísica y Astronomía, CONICET Universidad Nacional de San Juan, Av. Ignacio de la Roza 590 (O), J5402DCS, Rivadavia, San Juan, Argentina
         \and
        Universidad Nacional de Córdoba (UNC). Observatorio Astronómico de Córdoba (OAC). Laprida 854, Córdoba X5000BGR, Argentina.
        \and
         Departamento de Física, Universidad Técnica Federico Santa María, Avenida Vicuña Mackenna 3939, San Joaquín, Santiago, Chile
         \and
         Departamento de Física, Universidad Técnica Federico Santa María, Casilla 110-V, Avda. España 1680, Valparaíso, Chile
         \and
         Instituto de Física, Pontificia Universidad Católica de Valparaíso, Casilla 4950, Valparaíso, Chile
         \and
         Universidad Andres Bello, Facultad de Ciencias Exactas, Departamento de Fisica y Astronomia, Instituto de Astrofisica, Fernandez Concha 700, Las Condes, Santiago RM, Chile.
        \and
        CONICET. Instituto de Astronomía Teórica y Experimental (IATE). Laprida 854, Córdoba X5000BGR, Argentina.  
            }

   \date{Received --; accepted --}

\abstract
{
We investigate the large-scale conformity signal and its connection with the large-scale structure (LSS), focusing on low-mass central galaxies with stellar masses between $10^{9.5}$ and $10^{10}~h^{-1}$ M$_{\odot}$, after excluding galaxies that were satellites in the past. We measure the signal across different environments identified with the DisPerSE algorithm, including group outskirts, cluster outskirts, filaments, filament outskirts, and other environments (voids and walls) in order to assess how each population contributes to the large-scale conformity signal.

We use the galaxy catalogs from the IllustrisTNG300 simulation and the semi-analytic model MDPL2-SAG. In the semi-analytic model SAG, galaxies embedded in filamentary regions emerge as the primary contributors to the large-scale conformity signal, closely reproducing its overall shape and amplitude. In TNG300, galaxies located in filamentary regions also exhibit a coherent, albeit weaker, conformity signal once galaxies that were satellites in the past are excluded.

We also examine whether the conformity signal depends on the adopted filament definition. We find no substantial differences when varying the filament thickness or considering galaxies in overlapping regions. Instead, the signal remains strong and largely insensitive to these differences. We then investigate the dependence of the conformity signal on filament linear density, length, and mass. The signal is enhanced in high-linear density, short, and high-mass filaments ($M_F \geq 10^{13.39}~h^{-1}M_{\odot}$), but remains present in low-linear density, long, and low-mass filaments, although with a slightly lower amplitude.

Finally, we find that low-mass central galaxies located in close proximity ($\leq 0.1 \mathrm{Mpc}~h^{-1}$) to interacting massive systems show an amplified conformity signal extending up to $4~h^{-1}$Mpc. 

Overall, our results indicate that filamentary regions, including their immediate surroundings, and the tidal forces from neighboring massive systems play a fundamental role in shaping the large-scale conformity signal.
}


\titlerunning{The role of filaments in the large-scale conformity signal}
\authorrunning{Daniela Palma et al.}
\maketitle


\section{Introduction} \label{sec:intro}

The two-halo conformity (or large-scale conformity) refers to the correlation between the star-formation activity of central galaxies and that of their neighbors over megaparsec scales. This phenomenon, particularly prominent in the low-mass regime, has been extensively investigated in both observational (\citealt{Kauffmann_2013}; \citealt{Kauffmann_2015}; \citealt{Olsen_2023}) and theoretical (\citealt{Hearin_2015}, \citealt{Hearin_2016}, \citealt{Lacerna_2018}, \citealt{Lacerna_2022}, \citealt{Wang_2023}, \citealt{Ayromlou_2023}, \citealt{Palma_2025}) studies. Its physical origin, however, remains debated. 

Previous works have suggested that post-processed populations, such as backsplash or fly-by galaxies (hereafter former satellites), can significantly contribute to the signal, particularly in hydrodynamical models (e.g., \citealt{Wang_2023}, \citealt{Palma_2025}), while their impact appears smaller (or negligible) in semi-analytic models (e.g., \citealt{Ayromlou_2023}, \citealt{Palma_2025}). In this context, \cite{Lacerna_2022} showed that low-mass central galaxies with stellar masses between $10^{9.5}$ and $10^{10}~h^{-1}$ M$_{\odot}$ located within $5~h^{-1}$Mpc of groups and cluster centers drive the large-scale conformity signal. These extended zones host infalling and backsplash populations that may account for the strong conformity signal observed in these regions.

Recently, \citet{Palma_2025} demonstrated that the large-scale conformity signal of low-mass central galaxies can be quantitatively similar in two models with different prescriptions and resolutions: the hydrodynamical simulation IllustrisTNG300 (\citealt{Naiman_2018}; \citealt{Nelson_2018}; \citealt{Marinacci_2018}; \citealt{Pillepich_2018}; \citealt{Springel_2018}; \citealt{Nelson_2019}) and the semi-analytic model of galaxy formation SAG \citep{Cora_2018,Cora_2019}, applied to the dark matter (DM) only simulation MDPL2 \citep{Klypin_2016, Knebe_2018}. However, after the identification of former satellites, the signal in both models responds differently to the removal of former satellites. In particular, excluding post-processed galaxies strongly reduces the signal (between $75-85\%$ at $z=0$) in TNG300, whereas the impact in SAG is negligible.

A recent work by \citet{Lacerna_2025} showed that conformity in SAG using central galaxies with a fixed halo mass ($10^{11.6} \leq \rm M_{h}/h^{-1} M_{\odot} \leq 10^{11.8}$) is closely connected to galaxy assembly bias. This phenomenon is referred to the dependence of large-scale galaxy clustering on secondary halo properties beyond the halo mass (e.g., spin \citep{AMD_2020}, formation time \citep{Artale_2018}, or concentration \citep{Contreras_2019}). This result could point out to differences in halo growth histories that may imprint large-scale correlations in low-mass galaxy properties. The connection between the conformity signal and assembly bias have been proposed in previous works (e.g., \citealt{Hearin_2015}, \citealt{Paranjape_2015}, \citealt{Hearin_2016}, \citealt{AMD_2020}). These results suggest that galactic conformity is unlikely to arise from a single mechanism, but rather from the interplay between environmental processes and halo assembly histories.

For instance, \citet{Palma_2025} found systematic differences since $z < 1$ in dark matter, gas content and star-formation histories compared to systems far away from massive structures, evidencing that environmental effects extend well beyond the virial radii of massive halos. In fact, there is observational evidence that environmental effects extend up to five virial radii from the center of clusters \citep{Piraino-Cerda_2024}, providing evidence of pre-processing effects (\citealt{Fujita_2004}; \citealt{Wetzel_2013}; \citealt{Haines_2015}; \citealt{Pallero_2019}; \citealt{Hough_2023}) at these scales. Moreover, the evolution of gas content for quenched low-mass central galaxies near massive structures resembles the trends found for galaxies in filamentary infall regions in SAG \citep{Salerno_2022}, raising the possibility that filamentary environments may be central to the origin of the persistent conformity signal in the semi-analytic model. This interpretation is further supported by the fact that the vicinity regions of groups and clusters defined in megaparsecs inevitably encompass filamentary structures that act as bridges between virialized structures and the surrounding large-scale structure (LSS).

Filaments trace the cosmic web and provide the dynamical channels through which matter flows toward massive halos. Identified through topological algorithms such as DisPerSE \citep{Sousbie_2011a}, \citep{Sousbie_2011b} these structures connect groups and clusters across megaparsec scales. While numerous studies have explored galaxy properties in filaments and low-density regions (e.g., \citealt{Codis_2012}; \citealt{Welker_2014}; \citealt{Dubois_2014}; \citealt{Kuutma_2017}; \citealt{Laigle_2018}; \citealt{Kraljic_2018}; \citealt{Welker_2020}; \citealt{DGE_2021}; \citealt{Kraljic_2021}; \citealt{Song_2021}; \citealt{Malavasi_2022}; \citealt{Rodriguez_Medrano_2022}; \citealt{Hasan_2023}; \citealt{DGE_2023}; \citealt{Rodriguez_Medrano_2024}; \citealt{AMD_Rodriguez_2024}; \citealt{Stephenson_2025}; \citealt{Navdha_2025}; \citealt{Finn_2025}; \citealt{Zakharova_2025}; \citealt{Rodriguez_Medrano_2025}; \citealt{Benavides_2025}; \citealt{Barsanti_2025}), their specific role in the large-scale conformity signal remains unclear.

The impact of filamentary environments appears to depend strongly on stellar mass. \cite{OKane_2024} used a set of galaxies with stellar masses $\gtrsim 10^{10}$~M$_{\odot}$ from the SDSS (DR8, \citealt{SDSS_DR8}) to analyze the influence of filaments on the evolution of these galaxies. The results showed that filaments do not significantly contribute to the star-formation quenching at these stellar masses. Although galaxies in filaments exhibit lower star-formation activity than field galaxies, these differences vanish at fixed stellar mass and local galaxy density, suggesting that local density dominates over the filamentary environment for intermediate and massive galaxies. Whereas, open questions remain regarding the impact of these environments on low-mass galaxies, which are more susceptible to external influences and exhibit stronger correlations in star-formation with neighboring galaxies. Recently, \cite{Benavides_2025}, using galaxies from the IllustrisTNG50 with stellar masses in the range $10^{7}$–$10^{9} M_{\odot}$, showed that galaxies crossing filamentary regions in the past experience a phenomenon called cosmic web stripping, leading to a gas removal and an increased quenched fraction of galaxies in low-density regions at present. While massive galaxies appear largely insensitive to filamentary environments, and dwarf galaxies can be strongly affected by cosmic web stripping, the low-mass regime remains relatively unexplored.

The main goal of this work is to investigate the environmental impact of filamentary structures on the large-scale conformity signal produced by a population of low-mass central galaxies with stellar mass ranges between $10^{9.5}$ and $10^{10}~h^{-1}$ M$_{\odot}$, after removing post-processed populations in order to minimize the contribution of former satellites and better assess the role of the large-scale environment.
Since filamentary structures constitute a key component of the large-scale environment, they provide a natural framework for investigating the physical origin of the residual conformity signal.
By using two cosmological simulations, the semi-analytic model SAG and the hydrodynamical simulation TNG300, we classify galaxies into distinct large-scale environments of the cosmic web, namely group outskirts, cluster outskirts, filaments, and filament outskirts. By isolating low-mass central galaxies that have never been former satellite galaxies per environment, we aim to disentangle intrinsic large-scale environmental influences from post-processing that could be accounting for the large-scale conformity signal at $z = 0$.

The paper is organized as follows. In Sect. \ref{sec:data}, we present the galaxy samples taken from the hydrodynamical and semi-analytic models. Section \ref{sec:method} describes the methodology employed to extract filaments in each simulation using DisPerSE, as well as the measurement of the conformity signal. The results are presented in Sects. \ref{sec:results} and \ref{sec:gas_dm_sSFR}, where we analyze the conformity signal across large-scale environments, the role of filament properties, and the changes in the physical properties of galaxies residing in these environments. Finally, in Sect. \ref{sec:discussion}, we summarize and discuss our main findings.

\section{Data} \label{sec:data}

We study low-mass central galaxies with stellar masses in the range $10^{9.5}-$ $10^{10}~h^{-1}$ M$_{\odot}$
from the IllustrisTNG project 
and the MDPL2-SAG model. This stellar mass regime was selected because previous studies have shown that it exhibits a strong large scale conformity signal (\citealt{Lacerna_2022}; \citealt{Ayromlou_2023}; \citealt{Wang_2023}; \citealt{Palma_2025}), while remaining above the resolution limits of both simulations.

In \cite{Palma_2025}, a population of former satellite galaxies, including backsplash and fly-by objects was identified among low-mass central galaxies and shown to significantly contribute to the conformity signal in TNG300, while no significant contribution was found in MDPL2-SAG. Motivated by these results, we restrict our analysis to low-mass central galaxies that have never become satellites of another DM halo, in order to isolate the intrinsic conformity signal unrelated to pre-processing effects. We focus our analysis on $z=0$, where a residual conformity signal persists even after removing former satellites.

The galaxy samples were selected from the largest volume of the hydrodynamical IllustrisTNG simulation, IllustrisTNG300 (TNG300 hereafter), which has a box length of 205 $h^{-1}$Mpc, a dark matter particle mass of $m_{\rm DM} \sim 4\times~10^{7}~h^{-1}M_{\odot}$, and a baryonic mass resolution of $m{\rm b} \sim 7.5 \times 10^{6}~h^{-1} M_{\odot}$. The gravitational softening length is $\epsilon \sim 1.48$ kpc for the dark matter and stellar components, and adaptive for the gas component with a minimum of $\sim 0.37$ kpc \citep{Pillepich_2018}. Star formation follows the subgrid multiphase model of \cite{Springel_Hernquist_2003}, while stellar and AGN feedback are implemented following \cite{Pillepich_2018} and \cite{Weinberger_2017}. Galaxies at the lower and upper stellar mass limits of our sample ($10^{9.5}$--$10^{10}~h^{-1} M_{\odot}$) are resolved with approximately $4 \times 10^{2}$ and $1.3 \times 10^{3}$ baryonic particles, respectively, well above the resolution limit of the simulation. Therefore, the stellar mass cutoff adopted for our galaxy samples are resolved with a sufficient number of stellar particles to ensure that the derived star-formation rates are reliable. 

The galaxy catalog MDPL2-SAG (SAG hereafter) combines the latest version of the semi-analytic model of galaxy formation SAG with the cosmological DM MULTIDARK Planck 2 simulation MDPL2 which spans a box length of $1~h^{-1}$Gpc and has a dark matter particle mass of $m_{\rm DM} \sim 1.5\times 10^{9}~h^{-1}M_{\odot}$. SAG models galaxy evolution through physical prescriptions for radiative cooling, star formation, chemical enrichment, supernova and AGN feedback, as well as environmental processes such as ram-pressure stripping and tidal stripping. In the model, the mass of hot gas available to cool and form stars is initially determined by applying the cosmic baryon fraction to the host halo virial mass, and is subsequently updated according to the gas cooling rate and the amount of gas already converted into stars. Star formation proceeds through both a quiescent mode, regulated by the cold gas content of the disc, and a starburst mode triggered by mergers and disc instabilities \citep{Cora_2016, Cora_2018}. The evolution of the baryonic components is coupled to the properties and assembly history of the host dark matter halo, allowing the model to follow the growth and quenching of galaxies within a cosmological framework.

The full sample contains 37,535 and 6,243,661 low-mass central galaxies at $z$ = 0 in TNG300 and SAG, respectively.

\section{Methods} \label{sec:method}

Despite the substantial differences in numerical resolution and galaxy formation prescriptions between the two models, the overall conformity signal at $z =0$ measured for low-mass central galaxies is quantitatively similar in both simulations (see black solid lines in Figs. 7 and 8 of \citealt{Palma_2025}), though its physical interpretation differs. In TNG300, the signal is partly explained by a population of former satellites, highlighting the susceptibility of low-mass systems to environmental influences. In contrast, a significant conformity signal remains in SAG even after removing former satellites, suggesting that additional large-scale environmental processes could contribute to the signal. This motivates a more detailed characterization of the large-scale environment in which these galaxies reside.

In previous works, such as \cite{Lacerna_2022} and \cite{Palma_2025}, the neighborhood of massive structures used to characterize their large-scale environment was defined using a fixed distance of 5 $h^{-1}$ Mpc from their centers. However, this approach may introduce biases, particularly in the characterization of group outskirts, as it covers relatively larger areas for groups compared to clusters. 

\subsection{Cosmic web description}
To identify the components of the cosmic web, we employ the publicly available code DisPerSE (\citealt{Sousbie_2011a}; \citealt{Sousbie_2011b}), a widely validated tool in the literature (e.g., \citealt{Perez_2024}; \citealt{Luber_2025}; \citealt{Rodriguez_AMD_2025}; \citealt{Wang_2025}; \citealt{Bahe_2025}; \citealt{Gallagher_2026}) that identifies multiscale structures based on discrete Morse theory. This framework uses the mathematical properties relating the topology and geometry of the density field to characterize the cosmic web by reconstructing the density field from a discrete set of tracers, such as DM halos or galaxies, using the Delaunay Tessellation Field Estimator (DTFE).
The algorithm then identifies critical points where the density gradient vanishes, classifying them as maxima, minima, or saddle points. Within the Morse theory framework, filaments are defined as the sets of integral lines (segments) connecting density maxima and saddle points.

To mitigate the identification of spurious structures or numerical noise arising from the discrete sampling of the density field, DisPerSE incorporates the concept of topological persistence. Persistence quantifies the robustness of pairs of critical points based on their density contrast, allowing the application of a threshold that removes low-significance features. This procedure ensures a physically stable identification of the filamentary network. For further details on the methodology, we refer the reader to \cite{Sousbie_2011a}. 
In this work, we adopted the parameter choices described in \cite{DGE_2024}, and \cite{DGE_2020} for the SAG and TNG300 simulations, respectively. For SAG, we traced the cosmic web using the most luminous galaxies ($M{\rm r} < -21$) in the full simulation box, adopting a $2\sigma$ persistence and one smoothing cycle. For TNG300, the cosmic web was reconstructed from galaxies with stellar masses $M \geq 10^9 M_{\odot}$, applying a $3\sigma$ persistence to the DTFE density field with one smoothing cycle.

The properties of the filament population identified in each simulation are summarized in Table \ref{tab:properties_filament_population}. We report the minimum, median, and maximum values of the filament luminosity (L$_{\rm fil}$), length, and linear density ($\mu_{\rm fil} \equiv$ L$_{\rm fil}/$length), considering only filaments associated with at least one central galaxy in our sample. The median values are later adopted as thresholds to divide the filament population into short and long, and low-linear and high-linear density subsamples.

\begin{table}[ht!]
 \caption{Properties of the filament population identified in the SAG and TNG300 simulations.}
\label{tab:properties_filament_population}
\begin{center}
 \renewcommand{\arraystretch}{1.4}
 \begin{tabular}{l|c|c}
Properties & SAG & TNG300\\
\hline 
L$_{\rm fil}$ ($10^{10}$ L$_{\odot}$)& 16.5 (1.9 $-$ 330) & 13.8 (1.9 $-$ 129) \\
Length (Mpc) & 21.0 (0.1 $-$ 171) & 13.1 (1.8 $-$ 106) \\
$\mu_{\rm fil}$ ($10^{9}$ L$_{\odot}$/Mpc) & 7.7 (0.3 $-$ 680) & 9.9 (0.6 $-$ 146) \\
 \end{tabular}
\tablefoot{Values are reported as median (minimum $-$ maximum) for luminosity (L$_{\rm fil}$), length, and linear density ($\mu_{\rm fil}$).}
\end{center}
\end{table}

Despite the different tracer populations and persistence thresholds used to identify filaments with DisPerSE in each simulation, the filament populations exhibit similar linear densities. The median filament luminosities and lengths are larger in SAG than in TNG300. These differences are expected given the different simulation volumes, tracer populations, and the persistence thresholds adopted to reconstruct each filament catalog. In particular, the larger median filament luminosity and length in SAG likely arise because the adopted tracer population (luminous galaxies with $M_{\rm r} < -21$), preferentially traces the most luminous filamentary structures, despite the lower overall number density of tracers compared to TNG300. This is also reflected in the larger maximum linear density measured in SAG. Nevertheless, as we show throughout this work, the main trends of the conformity signal persist within filamentary regions in both models.

Based on \citet{Galarraga-Espinoza_2023}, we define five regions: \textit{i)} galaxy group outskirts, \textit{ii)} galaxy cluster outskirts, \textit{iii)} filaments, \textit{iv)} outskirts of filaments, and v) other environments. 
We applied a mass-based criterion to distinguish between galaxy groups and clusters: systems with $M_{200} \leq~10^{14}~M_{\odot}$ were classified as groups, while more massive systems ($M_{200} >~10^{14}~M_{\odot}$) were classified as clusters. The outskirts of galaxy groups and clusters are defined as regions between 1 - 3 virial radii of each system. Galaxies embedded in filaments are located at perpendicular distances $\leq 1$ $h^{-1}$ Mpc, while filament outskirts correspond to distances between 1 and 2 $h^{-1}$ Mpc. Galaxies that do not belong to any of these regions are classified as residing in other environments. 
We do not consider galaxies located in nodes, that is, within galaxy groups or clusters, as our goal is to explore the connection between the large-scale conformity signal, produced by low-mass central galaxies located outside groups and clusters, and the cosmic web. By excluding nodes, we minimize the impact of local environmental processes associated with massive halos, allowing us to focus on the role of the LSS traced by filaments and other low-density environments\footnote{Previous work has shown that large-scale conformity signal is driven by low-mass central galaxies in the outskirts of massive structures (\citealt{Lacerna_2022}).}.

Table \ref{tab:gals_TNG_SAG_in_diff_environments} presents the distribution of low-mass central galaxies with stellar masses between $10^{9.5}-$ $10^{10}~h^{-1}$ M$_{\odot}$ across the different environments. We split the galaxies according to their star formation activity following the same specific star formation rate (sSFR) cut used in previous works (\citealt{Lacerna_2022}; \citealt{Palma_2025}; \citealt{Lacerna_2025}), that is, star-forming galaxies satisfy sSFR $>~10^{-10.5}~h~yr^{-1}$, otherwise they are classified as quenched galaxies. This value is based on the results of \cite{Cora_2018}, where it was found to reproduce well the bimodality of galaxies within SAG. For consistency, we keep the same sSFR cut in TNG300\footnote{We tested the cut: sSFR $\sim 10^{-11}~yr^{-1}$, and found no significant changes in the main properties of both population of galaxies. However, this threshold yields a reduced sample of quenched galaxies in TNG300, which gives noisier results.}. We emphasize that the sample of central galaxies used in this study was traced back in time to ensure that all objects remained centrals throughout their evolution; therefore, no backsplash galaxies are included in our sample. This selection is intended to minimize environmental contamination in the measurement of the conformity signal.

\begin{table*}
 \caption{Number of quenched (Q) and star-forming (SF) low-mass central galaxies with log$_{10}(M_{\star})= [9.5, 10] ~h^{-1}M_{\odot}$ in the SAG and TNG300 models distributed in different large-scale environments.
 \label{tab:gals_TNG_SAG_in_diff_environments}}
\begin{center}
 \renewcommand{\arraystretch}{1.6}
 \begin{tabular}{l|c|c|c|c||c|c|c|c}
 \hline
Environment & Q gals in SAG & \% & SF gals in SAG & \%  & Q gals in TNG & \% & SF gals in TNG & \% \\
\hline \hline
Group outskirts & 28976 & 5.4 & 512019 & 94.6 & 45 & 14.7 & 261 & 85.3\\
Cluster outskirts & 16907 & 8.8 & 175833 & 91.2 & 56 & 13.9 & 348 & 86.1\\
Filaments & 2047 & 1.7 & 120094 & 98.3 & 427 & 9.6 & 4019 & 90.4\\
Filaments outskirts & 2963 & 1.3 & 232344 & 98.7 & 452 & 9.4 & 4371 & 90.6 \\
Other environment & 18237 & 0.4 & 5134241 & 99.6 & 2135 & 7.8 & 25421 & 92.2 \\
\hline
Total & 69130 & 1.1 & 6174531 & 98.9 & 3115 & 8.3 & 34420 & 91.7\\ 
 \hline 
 \end{tabular}
\end{center}
\end{table*}

\begin{figure*}[htbp]
  \centering
  \begin{subfigure}[b]{0.47\textwidth}
    \centering
    \includegraphics[width=\textwidth]{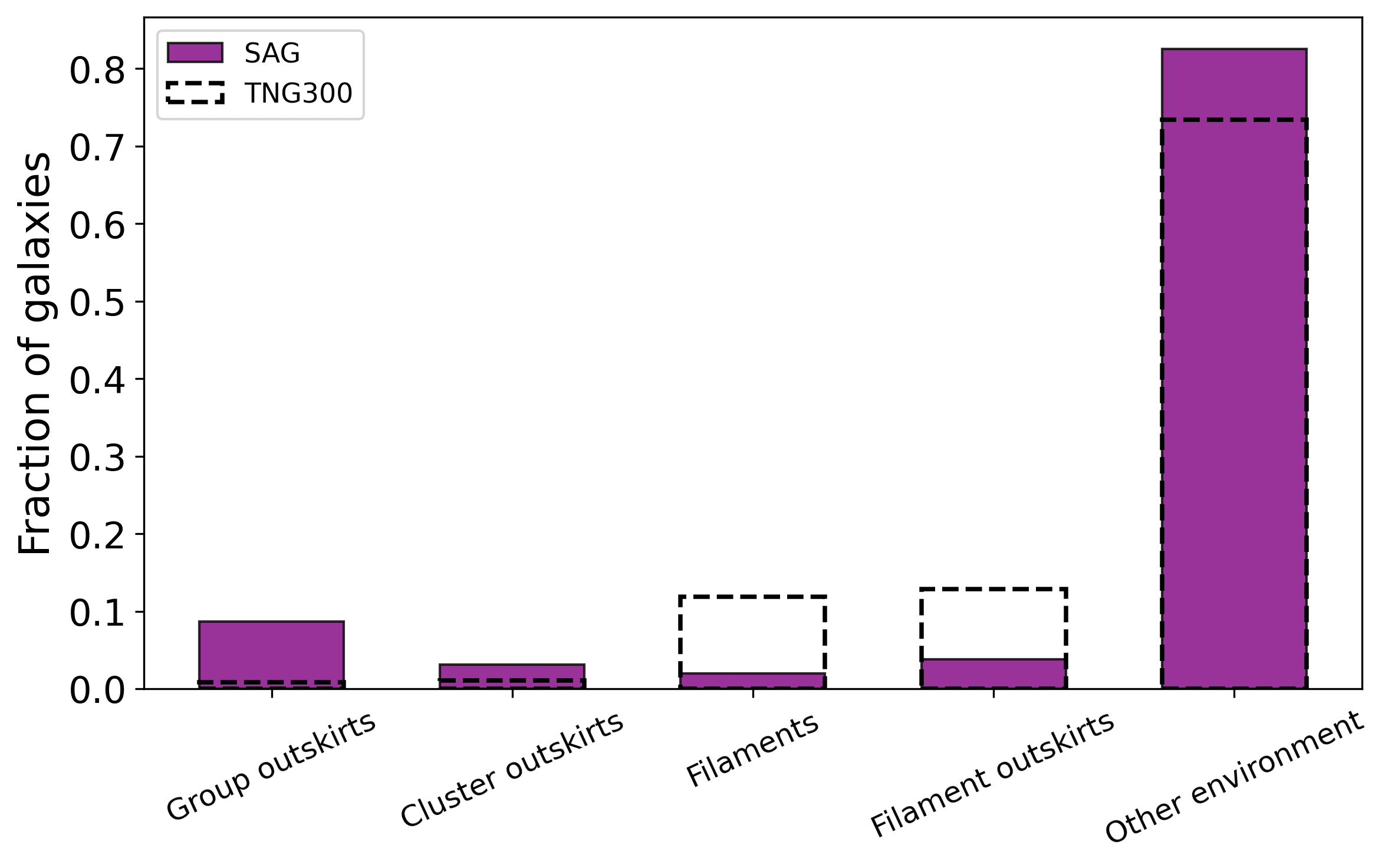}
  \end{subfigure}
  \begin{subfigure}[b]{0.474\textwidth}
    \centering
    \includegraphics[width=\textwidth]{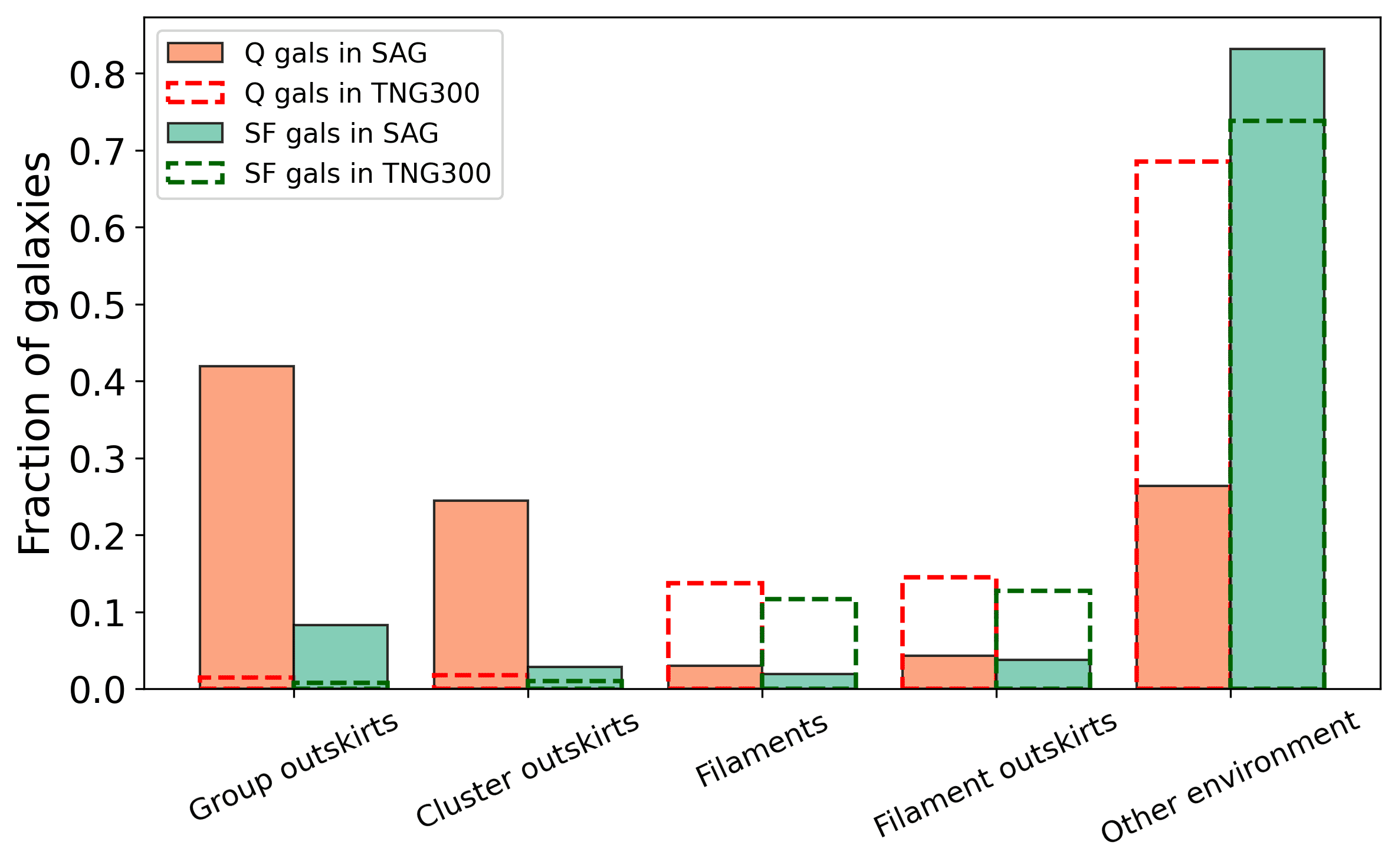}
  \end{subfigure}
  \caption{Left panels: Distribution of low-mass central galaxies with stellar masses in the range $9.5 \leq \log(M_*/$ $h^{-1}M_{\odot}$) $\leq 10$ from SAG (purple) and TNG300 (black dashed line), residing in different environments of the cosmic web. 
  Right panels: Same as in the left panel, but separating galaxies according to their star formation activity (star-forming in green and quenched galaxies in peach).}
  \label{fig:hist_gals_in_diff_environments}
\end{figure*}

The left panel of Fig. \ref{fig:hist_gals_in_diff_environments} shows the distribution of low-mass central galaxies across the different environments for SAG (filled purple) and TNG300 (black dashed line). In both models, most galaxies reside in other environments, such as voids or walls, according to the classification of \cite{Galarraga-Espinoza_2023}. This result is consistent with previous studies showing that these environments occupy the largest fraction of the simulation volume (e.g., using DisPerSE \citealt{Rodriguez_2025}; see also Fig. 8 in \citealt{Cautun_2014} based on NEXUS algorithm; and Tab. 1 in \citealt{Veena_2019}). Estimates of the volumetric fraction associated with filaments typically correspond to only a few percent of the total volume, depending on the parameters adopted to identify the structures. For instance, varying the persistence threshold and filament thickness in DisPerSE leads to only minor changes in the recovered filament volume fraction, but does not significantly alter the overall picture in which most of the volume is occupied by lower-density environments. Therefore, the predominance of galaxies in other environments is broadly consistent with the large volume fraction associated with these regions. Moreover, as previously shown by \cite{Palma_2025}, these galaxies are predominantly star-forming, which is consistent with their preferential position in low-density environments. This contrasts with galaxies in the same mass regime that reside in denser environments, such as galaxy clusters and groups, where local environmental effects can act more efficiently to suppress star formation.

The fraction of galaxies in filamentary regions (filaments and filament outskirts) is approximately four times higher in TNG300 than in SAG. Conversely, SAG exhibits a larger fraction of galaxies in the outskirts of groups and clusters than in TNG300.
These differences may partly reflect the distinct tracer populations adopted to reconstruct the cosmic web in each simulation, which affect both the abundance and spatial extent of the identified filamentary structures.

When separating galaxies by star-formation activity (right panel of Fig. \ref{fig:hist_gals_in_diff_environments}), star-forming galaxies (in filled green for SAG and green dashed line for TNG300) predominantly inhabit other environments in both models (over 70\% in TNG300 and $\sim$80\% in SAG), consistent with their dominance in the overall sample of low-mass central galaxies\footnote{As shown in Fig. 8 of \citealt{Rodriguez_Puebla_2015}, low-mass halos are more likely to host star-forming galaxies than high-mass halos.}. In contrast, quenched galaxies (in filled peach for SAG and red dashed line for TNG300) in the semi-analytic model are preferentially located in the outskirts of groups  ($\sim$40\%) and clusters ($\sim$25\%), while only a small fraction reside in filamentary regions ($<$7\%). In TNG300, however, quenched galaxies resemble the star-forming population, with the majority ($\sim$70\%) located in other environments.

\subsection{Conformity signal measurement} \label{sec:conformity_method}

After identifying galaxies in different large-scale environments, we estimate how each of them contributes to the large-scale conformity signal. As discussed by \cite{Palma_2025}, part of this signal may be driven by galaxies residing in filamentary regions. Their behavior, particularly in terms of gas mass evolution, resembles the trends reported by \cite{Salerno_2022}, who studied the gas content of quenched galaxies in filamentary infall regions using the SAG model. In both studies, galaxies in the same stellar mass range exhibit a decline in gas content since $z<1$, suggesting a common evolutionary trend.

The conformity signal is quantified by measuring the correlation between the sSFR of low-mass central galaxies and that of their neighbors out to megaparsec scales. Here, we focus on studying the signal at $z=0$. We estimated the quenched fraction $f_{\rm Q}$ around each low-mass central galaxy at radial separations out to 10 $h^{-1}$ Mpc.

We then calculated the mean quenched fraction of neighboring galaxies with stellar masses above $10^{9} h^{-1}M_{\odot}$ as a function of distance from quenched, $f_{\rm Q}^{\rm Q}(r_{\rm cen})$, and star-forming, $f_{\rm Q}^{\rm SF}(r_{\rm cen})$, central galaxies in each environment. The associated error in each distance bin was estimated using 1000 bootstrap realizations for both simulations. Finally, we measured the difference between the mean quenched fractions around quenched and star-forming central galaxies, $\Delta f_{\rm Q}(r_{\rm cen})$, as follows

\begin{center}
$\Delta f_{\rm Q} (r_{\rm cen})=f_{\rm Q}^{\rm Q}(r_{\rm cen})-f_{\rm Q}^{\rm SF}(r_{\rm cen})$,
\end{center}
for group and cluster outskirts, filaments, filament outskirts, and other environments in both simulations to assess the dependence of each environment on the large-scale conformity signal.

\section{Results} \label{sec:results}
We first investigate how different large-scale environments contribute to the correlation between low-mass central galaxies and their neighbors. We then explore whether the properties of filamentary structures, such as length, linear density, thickness, and superposition regions influence the signal.

\subsection{Conformity signal across cosmic web environments}
\begin{figure*}[htbp]
  \centering
    \includegraphics[width=\textwidth]{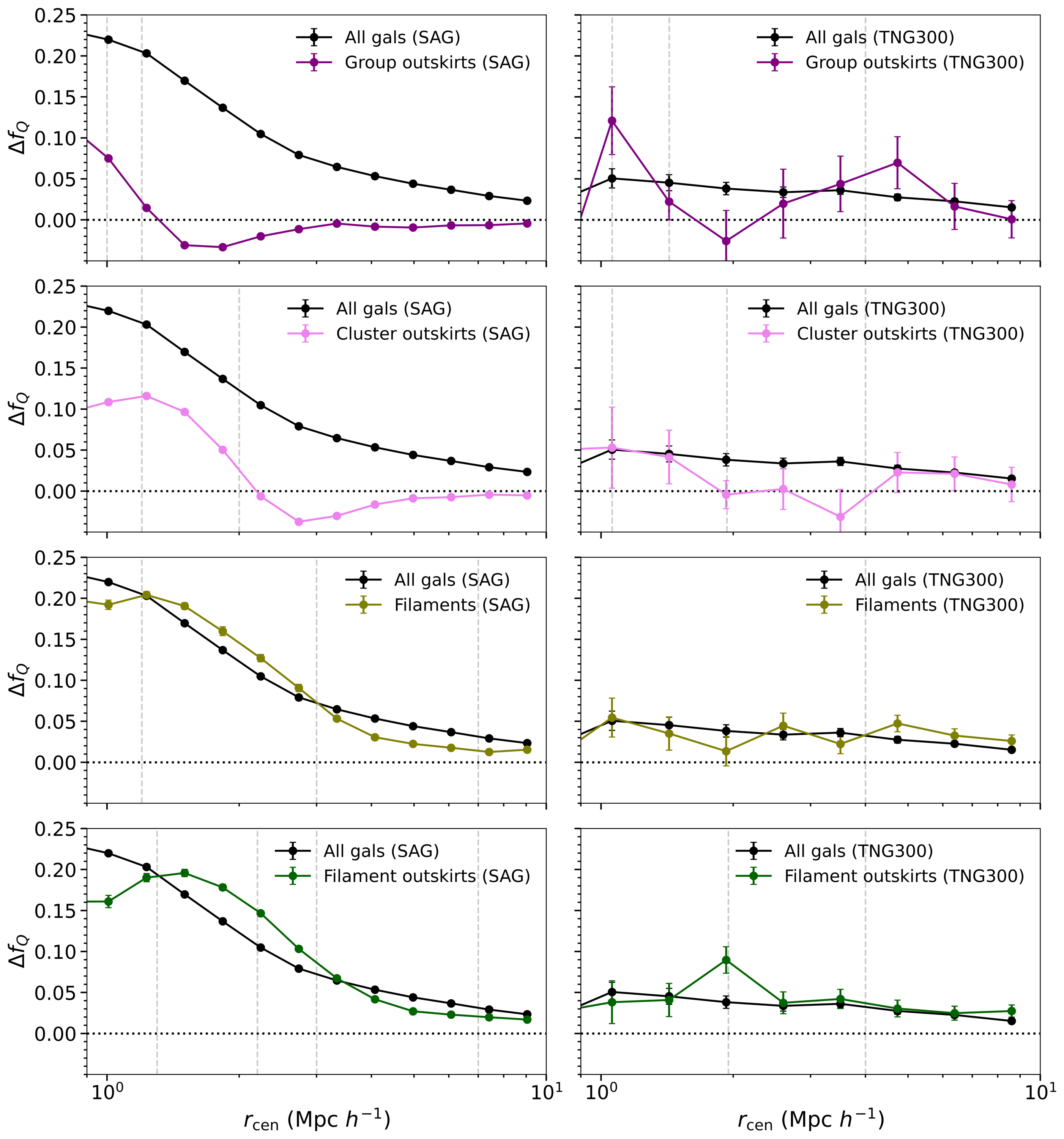}
  \caption{Conformity signal for all low-mass central galaxies in each model after removing backsplash objects (black solid line), compared with the signal obtained after classifying the galaxies per environment (colored lines). From top to bottom: Group outskirts (purple), cluster outskirts (violet), filaments (olive), and filament outskirts (dark green). Gray dashed vertical lines highlight transition scales discussed in Sect. 4.1. Left panels: semi-analytic model, SAG. Right panels: hydrodynamical model, TNG300.}
  \label{fig:confin_diff_environments}
\end{figure*}

Figure \ref{fig:confin_diff_environments} shows the two-halo conformity measurements, $\Delta f_{\rm Q}$, in SAG (left) and TNG300 (right) for low-mass central galaxies with stellar masses in the range $10^{9.5} - 10^{10}~h^{-1}~M_{\odot}$. The black solid lines correspond to all low-mass central galaxies, independent of their environment. These values were previously reported in \citet[][gray dashed lines in their Figs. 7 and 8]{Palma_2025}. The substantial differences in the signal amplitude between the models\footnote{The large-scale conformity signal measured for the full sample of low-mass central galaxies is quantitatively similar in both models, even with the differences in numerical resolution and prescriptions of each model. The differences in the signal appeared after the former satellite population was removed.}, motivates the present analysis of the role of filaments in shaping the large-scale conformity signal. In particular, the semi-analytic model remains of interest because it still exhibits a strong conformity signal even after removing former satellite galaxies.

The colored solid lines represent the signal produced by galaxies across different large-scale environments. From top to bottom, the panels show the contribution to the signal of galaxies located in group outskirts (purple), cluster outskirts (violet), filaments (olive), and filament outskirts (dark green). We show the results obtained analyzing the signal in other environments in Appendix \ref{sec:appendixA}. For completeness and consistency, we report the quenched fraction for each population in Appendix \ref{sec:appendixB}.

Looking at the left panels of Fig. \ref{fig:confin_diff_environments}, we examine how different populations of low-mass central galaxies in SAG contribute to the large-scale conformity signal. Galaxies located in the outskirts of groups and clusters show a signal primarily at small separations. The signal measured for centrals in group outskirts reaches an amplitude corresponding to 34\% of the total signal at $\sim 1~h^{-1}$~Mpc and declines rapidly beyond $\sim 1.2h^{-1}$~Mpc. Galaxies in cluster outskirts reach $56\%$ of the total signal amplitude at $\sim 1.2~h^{-1}$~Mpc, decreasing toward larger distances (36\% at $\sim 2.0~h^{-1}$~Mpc). Beyond these distances, no further contribution is detected. These transitions are highlighted in the figure as gray dashed vertical lines.
This behavior confirms that defining outskirts in terms of their virial radii provides a controlled characterization of their environmental impact. 

Galaxies residing in filamentary regions (filaments and filament outskirts) closely reproduce the shape and amplitude of the total signal. At separations below $1.2~h^{-1}$Mpc, the conformity signal measured for galaxies in filaments and filament outskirts reaches amplitudes of $99\%$ and $93\%$ of the total signal, respectively. 
At intermediate separations (between $\sim 1.2$ and $3 ~h^{-1}$~Mpc), the signal in these environments exceeds the amplitude of the full sample, reaching a peak excess of $\sim 20\%$ for filaments and $\sim 40\%$ for filament outskirts around $2.2~h^{-1}$Mpc. This indicate that filamentary regions host the strongest conformity signal at these scales. The remarkably similar behavior observed for galaxies within filaments and in their immediate surroundings further suggests that the conformity signal is not confined to the filament spine itself, but extends throughout the surrounding filamentary environment. This behavior is expected because the total signal represents the average over all environments, including populations that show weaker or negligible conformity, thereby reducing the overall amplitude. Beyond $\sim 3~h^{-1}$Mpc, the contribution of filament regions gradually declines, although the signal remains detectable up to $\sim 7~h^{-1}$~Mpc. Despite their smaller number, galaxies in filamentary environments appear to play a major role in shaping the conformity signal in the semi-analytic model.

The right panels of Fig. \ref{fig:confin_diff_environments} show the corresponding measurements for TNG300. Due to the smaller galaxy sample size compared to SAG (see Table \ref{tab:gals_TNG_SAG_in_diff_environments} for details), the associated uncertainties are larger. For this model, we focus on distance bins where the statistical significance is sufficient (i.e., $|\Delta f_{\rm Q}|~\gtrsim~1\sigma_{\Delta f_{\rm Q}}$), and where conformity is expected based on previous results. Therefore, we restrict this analysis to separations up to $\sim 4~h^{-1}$Mpc, where it was shown \citep{Palma_2025} that after removing former satellites the conformity signal becomes weak at larger distances. Within this range, filamentary regions provide the only statistically significant contribution to the signal, supported by both $\Delta f_{\rm Q}$ and $f_{\rm Q}$ (see Fig. \ref{fig:fQ_in_diff_environments}) measurements. Galaxies in filamentary regions closely reproduce the global conformity signal over the full range of scales where it is detected, indicating that these environments sustain the conformity signal once the contribution from massive structures become negligible. In particular, galaxies in filament outskirts exhibit a local enhancement around $\sim 2~h^{-1}$Mpc in both models, a feature that is especially pronounced in TNG300. 
Although $\Delta f_{\rm Q}$ is affected by large uncertainties in galaxies located in group and cluster outskirts, the examination of $f_{\rm Q}$ (Fig. \ref{fig:fQ_in_diff_environments}) further corroborates that quenched fractions around quenched and star-forming central galaxies remain similar, at least at large distance bins, with no systematic separation indicative of conformity. At smallest separations ($\sim 1~h^{-1}$Mpc), however, a marginal signal is visible in both group and cluster outskirts. While the trend for group outskirts is consistently recovered in both $\Delta f_{\rm Q}$ and $f_{\rm Q}$, the corresponding signal for cluster outskirts remains comparable to the associated uncertainties. Overall, the behavior qualitatively resembles that observed in SAG. The signal is no longer detected beyond $\sim 1.4~h^{-1}$Mpc and $\sim 1.9~h^{-1}$Mpc in group and cluster outskirts, respectively.

Taken together, these results demonstrate that the conformity amplitude clearly depends on the large-scale environment in both models. Galaxies in the outskirts of groups and clusters contribute primarily at small separations, consistent with the localized influence of massive halos, an effect that is robustly detected in SAG and only marginally present in TNG300 given its more limited statistics. Beyond these scales, however, filamentary environments dominate the remaining conformity signal in both simulations, indicating that filaments and their immediate surroundings are capable of sustaining conformity beyond the direct influence of groups and clusters. Moreover, the light blue line in Fig. \ref{fig:confin_OE} implicitly represents the signal obtained after removing galaxies in the outskirts of groups and clusters and filamentary regions; the resulting decrease in amplitude reinforces the dominant role of these populations in shaping the global conformity signal (see Appendix \ref{sec:appendixA}).

Although former satellites dominate the overall conformity amplitude in TNG300 (\citealt{Wang_2023}; \citealt{Palma_2025}), galaxies in filamentary regions exhibit an intrinsic, albeit weaker, conformity signal that persists even after excluding former satellites. To further test the role of filamentary regions in shaping the conformity signal, we perform additional tests including former satellite galaxies to assess the robustness of the filamentary contribution against sample size effects. We consider all low-mass central galaxies at $z=0$, reducing the associated statistical uncertainties and taking advantage of the previously studied role of the former satellite population on the signal in the model. Appendix \ref{sec:appendixC} confirms that filamentary regions largely shape the global signal, previously found for this stellar mass regime (\citealt{Lacerna_2022}; \citealt{Palma_2025}) but without distinguishing between environments.

As a complementary test, we also evaluate the impact of removing galaxies by environment, following the approach of \cite{Lacerna_2022}, \cite{Ayromlou_2023}, \cite{Wang_2023}, and \cite{Palma_2025}. The results are presented in Appendix \ref{sec:appendixD} (see Fig. \ref{fig:confin_diff_environments_by_removing_per_pop}). We find that removing galaxies located in filamentary regions produces negligible changes in the total conformity signal in SAG. For TNG300 instead, the mean values exhibit a decreasing behavior, although within the uncertainties. On the contrary, removing galaxies in other environments (see gray dashed lines in Fig. \ref{fig:confin_OE}) leads to a noticeable decrease in its amplitude in both models. The signal vanishes in SAG beyond $\sim 3~h^{-1}$Mpc in SAG, and it keeps qualitatively similar to the global signal in TNG300. This behavior primarily reflects the relative size of the galaxy samples rather than reflecting intrinsic physical differences alone between environments. In particular, galaxies in filamentary regions represent a small fraction of the total population $\sim 5\%$ in SAG vs $\sim 25\%$ in TNG300, while galaxies in other environments dominate the statistics despite exhibiting a weak intrinsic conformity signal. 

\subsection{Dependence of the conformity signal on filament properties}
Since the focus of this work is on the role of filamentary structures in the large-scale conformity signal, we restrict our density analysis to the internal properties of filaments (i.e., their linear density and length) rather than to the general large-scale density field surrounding galaxies. The filament population identified in the simulations spans a range of lengths, masses, and linear densities (\citealt{DGE_2020}; \citealt{DGE_2021}; \citealt{DGE_2023}). These variations may lead to different interpretations of the results observed in the previous section. Furthermore, galaxies can be associated with more than one filament, and such overlapping regions may contribute to the strength of the measured signal previously observed in Fig. \ref{fig:confin_diff_environments}. 

In the previous section, we assigned galaxies to filaments based on their minimum perpendicular distance to the filament spine. This criterion was applied in both models. To analyze galaxies in overlapping regions, we measured how many galaxies lie within 1 $h^{-1}$Mpc of more than one filament spine, thus determining the fraction of galaxies associated with multiple filaments.

For SAG, we find that only 5\% of the galaxies are associated with more than one filament, whereas this fraction increases to $\sim$23\% in TNG300. Despite these differences, we checked that the conformity signal measured for galaxies associated with a single filament and for those located in overlapping filamentary regions remains consistent within the statistical uncertainties. In both models, the shape and amplitude of the signal are broadly consistent, suggesting that the signal is not driven by galaxies that reside in overlapping regions.
We also repeated the analysis using a maximum perpendicular distance of $0.5~h^{-1}$Mpc instead of $1~h^{-1}$Mpc to associate galaxies with filaments. The resulting conformity measurements remain consistent within the uncertainties over the full range of separations explored. Therefore, the presence and strength of the signal are robust against changes in the adopted filament thickness. Overall, these tests indicate that the role of filaments in shaping the conformity signal is robust against variations in the selected filament population.

In addition, we use the median values described in Table \ref{tab:properties_filament_population} to separate the filament population based on their length. That is, filaments with lengths smaller than 21.0 $h^{-1}$Mpc (13.1 $h^{-1}$Mpc) were selected as short filaments in SAG (TNG300); filaments with larger median lengths were selected as long filaments. Based on their linear density, filaments with linear density greater than $7.7 \times 10^{9} L_{\odot}/$Mpc ($9.9 \times 10^{9} L_{\odot}/$Mpc), were classified as high-linear density filaments in SAG (TNG300), while filaments with lower median values were classified as low-linear density filaments. For SAG, we obtain that 43\% vs 57\% of the galaxies are embedded in short vs long filaments; whereas 53\% vs 47\% are embedded in high-linear density vs low-linear density filaments. For TNG300, 45\% vs 55\% of the galaxies are embedded in short vs long filaments (very similar to the percentages found in SAG), whereas 49\% vs 51\% are embedded in high-linear density vs low-linear density filaments.

\begin{figure}[htbp]
  \centering
    \includegraphics[width=0.5\textwidth]{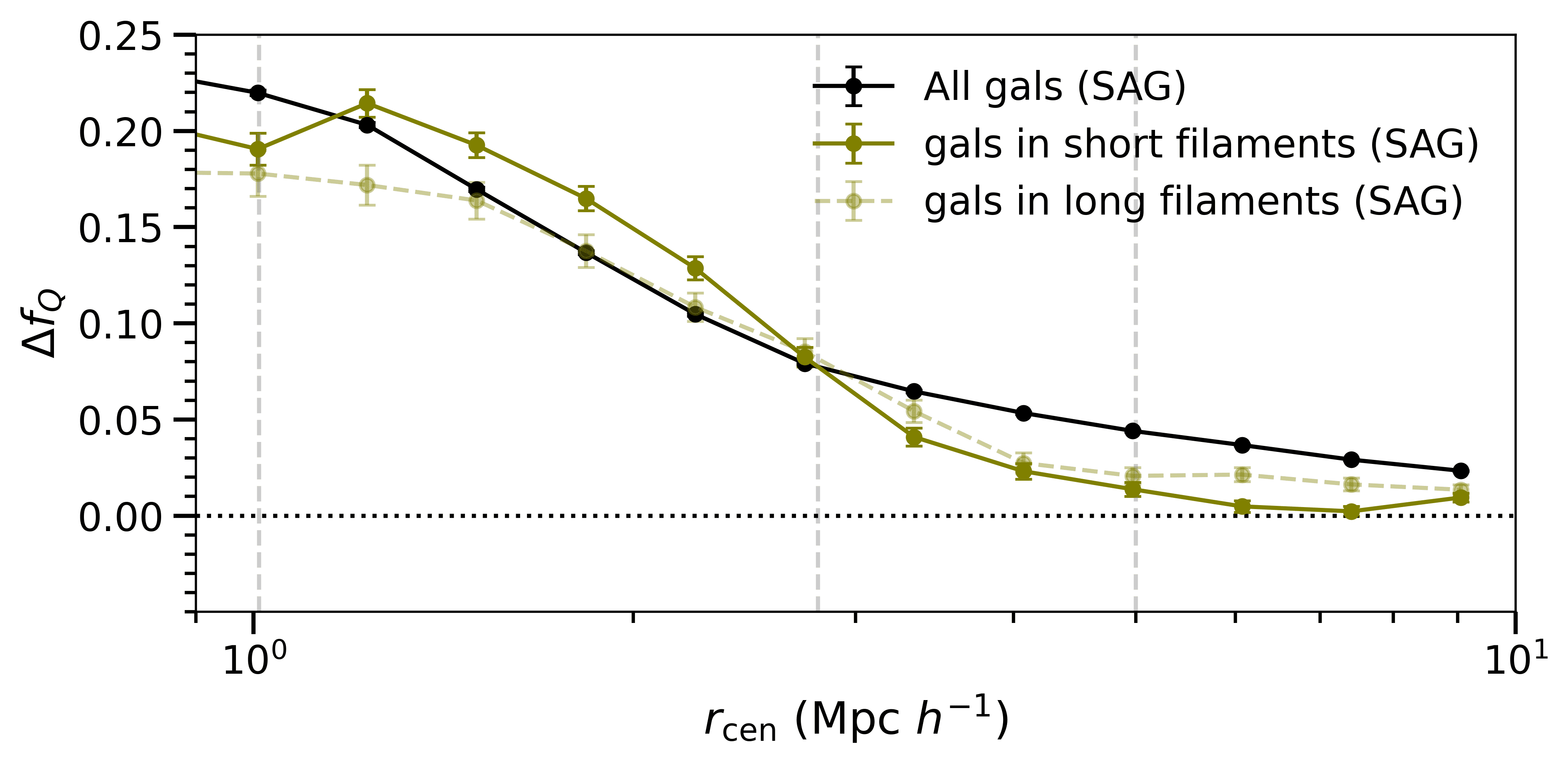}
  \caption{Conformity signal, $\Delta f_{\rm Q}$, for low-mass central galaxies in SAG, located in short (solid) and long (dashed) filaments.}
  \label{fig:conf_SAG_short_vs_large_fil}
\end{figure}

\begin{figure}[htbp]
  \centering
    \includegraphics[width=0.5\textwidth]{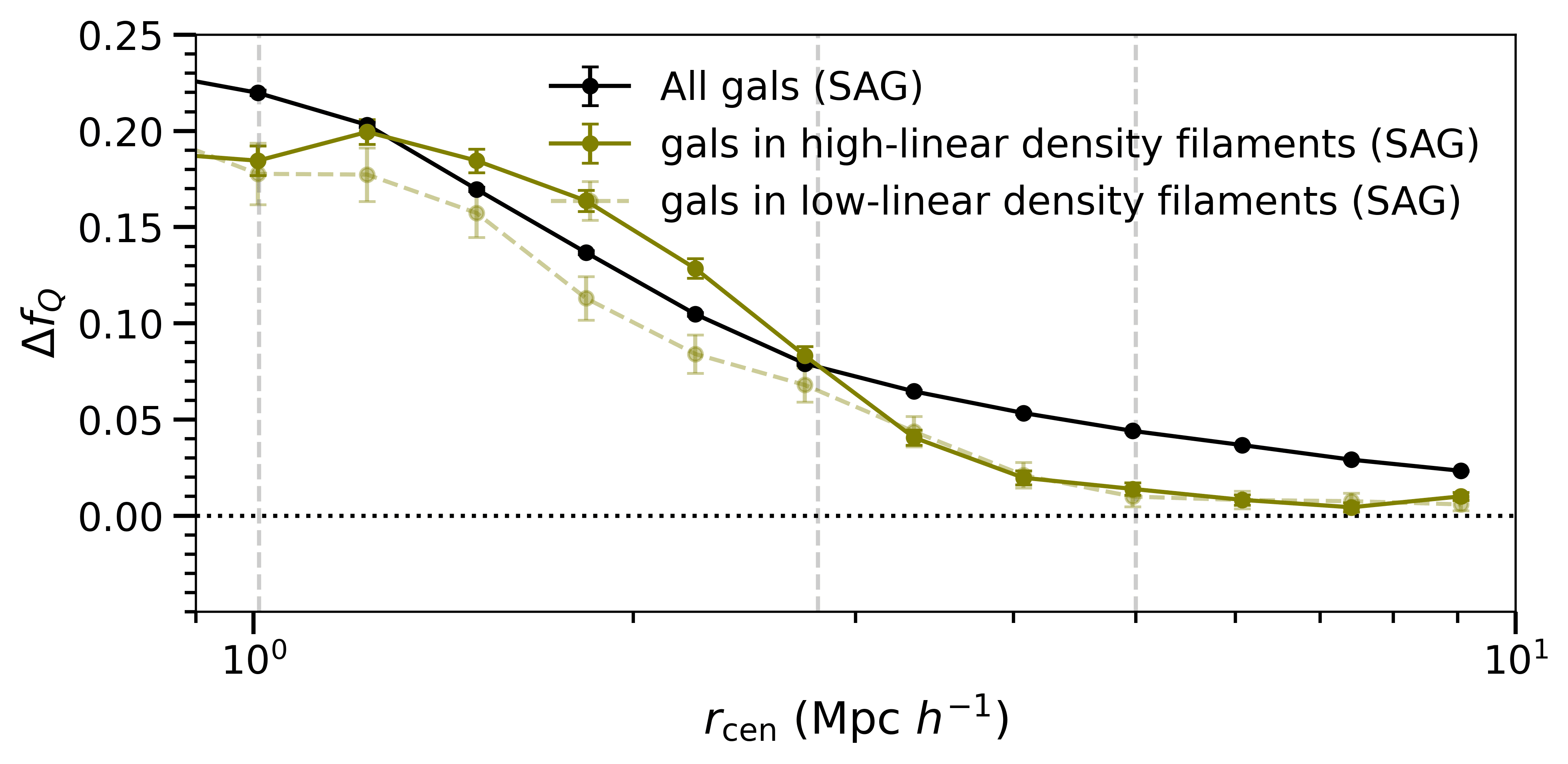}
  \caption{Conformity signal, $\Delta f_{\rm Q}$, for low-mass central galaxies in SAG, located in high-density (solid) and low-density (dashed) filaments.}
  \label{fig:conf_SAG_high_vs_low_density_fil}
\end{figure}

Figure \ref{fig:conf_SAG_short_vs_large_fil} shows the signal produced by low-mass central galaxies in SAG located in long (solid line) vs short (dashed line) filaments, and Fig. \ref{fig:conf_SAG_high_vs_low_density_fil} shows the conformity signal produced by galaxies in high-linear density (solid line) vs low-linear density (dashed line) filaments. For completeness, we tested these differences in TNG300 (see Appendix \ref{sec:appendixE}).

The figures show that galaxies residing in both long and short filaments, as well as in high- and low-linear density filaments, exhibit a clear conformity signal up to $\sim 5~h^{-1}$Mpc. The main differences arise at intermediate separations ($\sim 1.5 - 3~h^{-1}$Mpc), where short filaments (solid line in Fig. \ref{fig:conf_SAG_short_vs_large_fil}) and high-linear density filaments (solid line in Fig. \ref{fig:conf_SAG_high_vs_low_density_fil}) systematically exhibit enhanced signal amplitudes. At these scales, the signal measured for galaxies in short filaments reaches $\Delta f_{\rm Q} \sim 0.19-0.21$, exceeding that measured in long filaments by approximately $15 - 25\%$. Likewise, galaxies residing in high-linear density filaments exhibit amplitudes that are typically $\sim 10-20$\% larger than those measured in low-linear density filaments. Although both high- and low-linear density filaments display remarkably similar radial trends, the signal in high-linear density filaments systematically maintains a larger amplitude over these intermediate scales. These trends point out to a dependence of the conformity amplitude on the physical properties of the hosting filament, particularly its length and linear density.
Although galaxies in high-linear density filaments could in principle be more affected by environmental processes such as cosmic web stripping, the persistence of a clear conformity signal in low-linear density filaments indicate that this effect alone cannot explain the observed trends. Instead, the results suggest that filament properties regulate the amplitude of the conformity signal rather than determining its presence.
Furthermore, the stronger signal measured in short filaments provides additional support for this interpretation, since short filaments are known to be systematically denser than long filaments (\citealt{DGE_2020}).

We also test the dependence of the signal on filament mass. The results presented in Appendix \ref{sec:appendixF} show that the signal strengthens in high-mass filaments ($\geq~10^{13.39}~h^{-1}~M_{\odot}$), further supporting the role of filamentary structures in shaping the conformity signal and strengthening its connection with the large-scale environment.

Overall, these results confirm that filaments play an important role in shaping the large-scale conformity signal.

Thus, filamentary environments appear to sustain the residual large-scale conformity signal in the models once post-processed populations have been excluded. 

\section{Gas, Dark matter content, and sSFR} \label{sec:gas_dm_sSFR}

The environmental dependence identified in the large-scale conformity signal strongly suggests that density, and therefore the action of environmental processes in these regions, may play a key role in giving a physical explanation in shaping this effect. The enhanced susceptibility of low-mass galaxies to these mechanisms \citep{Chartab_2020} further supports this connection, and the distinct contributions of different large-scale environments to the signal reinforce this dependency (see Fig. \ref{fig:confin_diff_environments}). 

Environmental processes, independently of pre-processing effects, such as strangulation, can gradually deplete the gas reservoir of galaxies and suppress star formation. In addition, gas stripping induced by motion through filamentary environments (\citealt{Benavides_2025}), as well as tidal forces associated with massive structures, can contribute to the disruption of small halos (\citealt{Choque-Challapa_2019}). Thus, we examine the gas mass, sSFR, and dark matter content (in terms of M$_{200}$) distributions to analyze how these properties vary across environments for central galaxies. In particular, we investigate potential environmental signatures associated with these regions. 

\begin{figure}[htbp]
  \centering
    \includegraphics[width=0.5\textwidth]{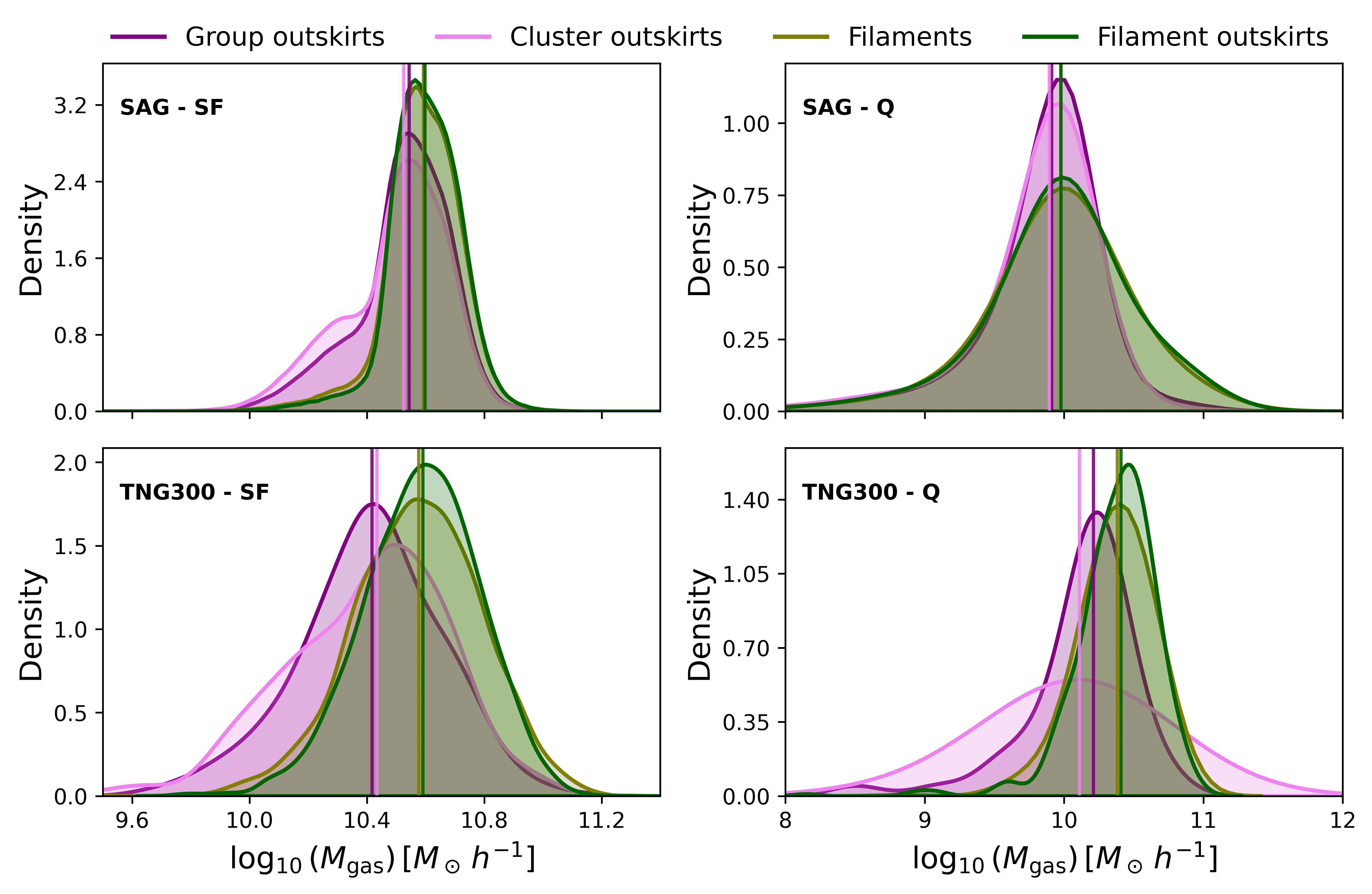}
  \caption{Probability distribution of gas mass for low-mass central galaxies across different environments. Star-forming and quenched galaxies are shown in left and right panels, respectively. The upper and lower panels correspond to SAG and TNG300, respectively. The vertical lines correspond to the median values colored per environment.}
  \label{fig:mgas_in_diff_environments}
\end{figure}

\begin{figure}[htbp]
  \centering
    \includegraphics[width=0.5\textwidth]{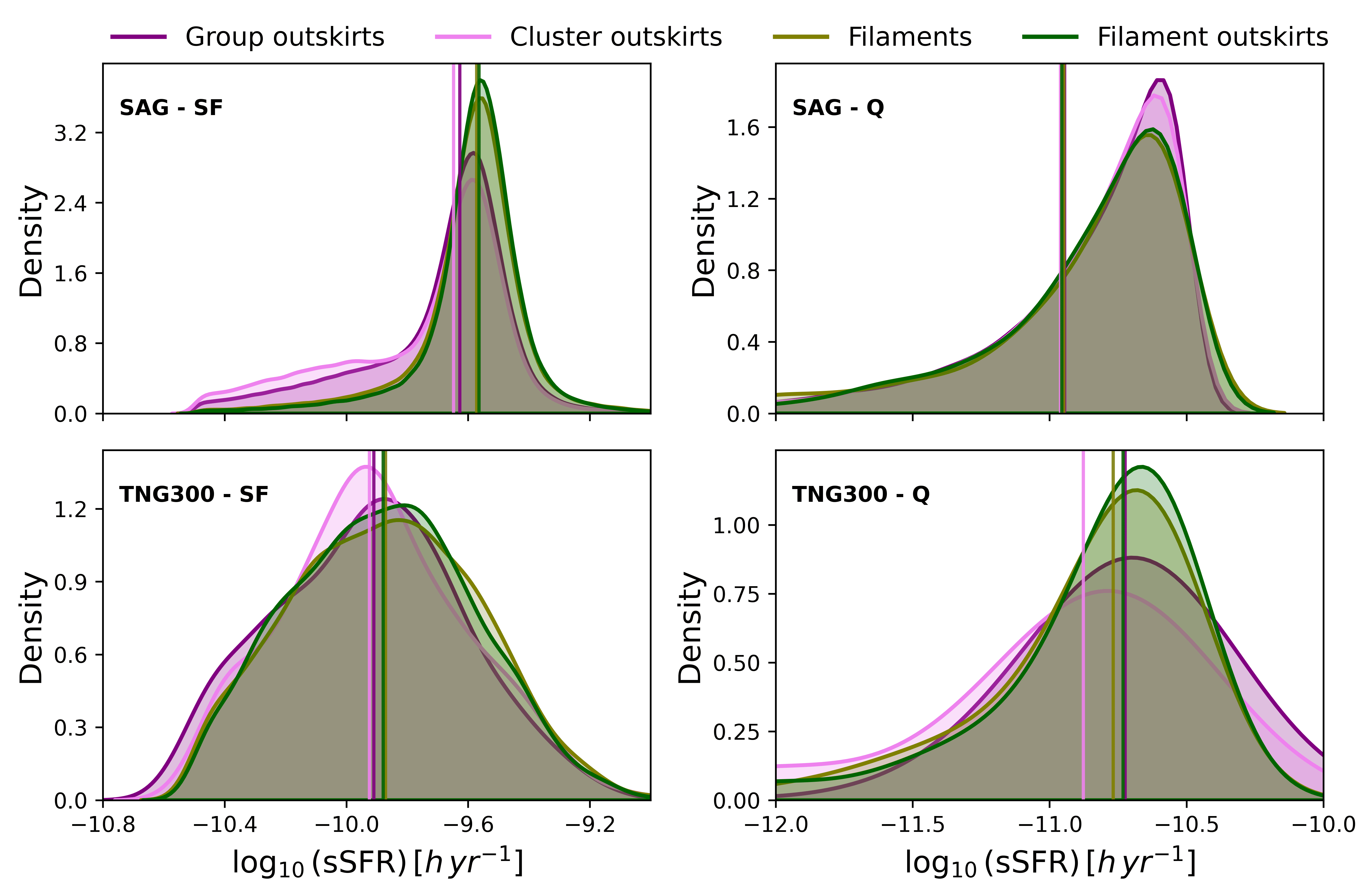}
  \caption{Probability distribution of sSFR for low-mass central galaxies across different environments. The panels and colors are designed in the same way as in Fig. \ref{fig:mgas_in_diff_environments}.}
  \label{fig:ssfr_in_diff_environments}
\end{figure}

\begin{figure}[htbp]
  \centering
    \includegraphics[width=0.5\textwidth]{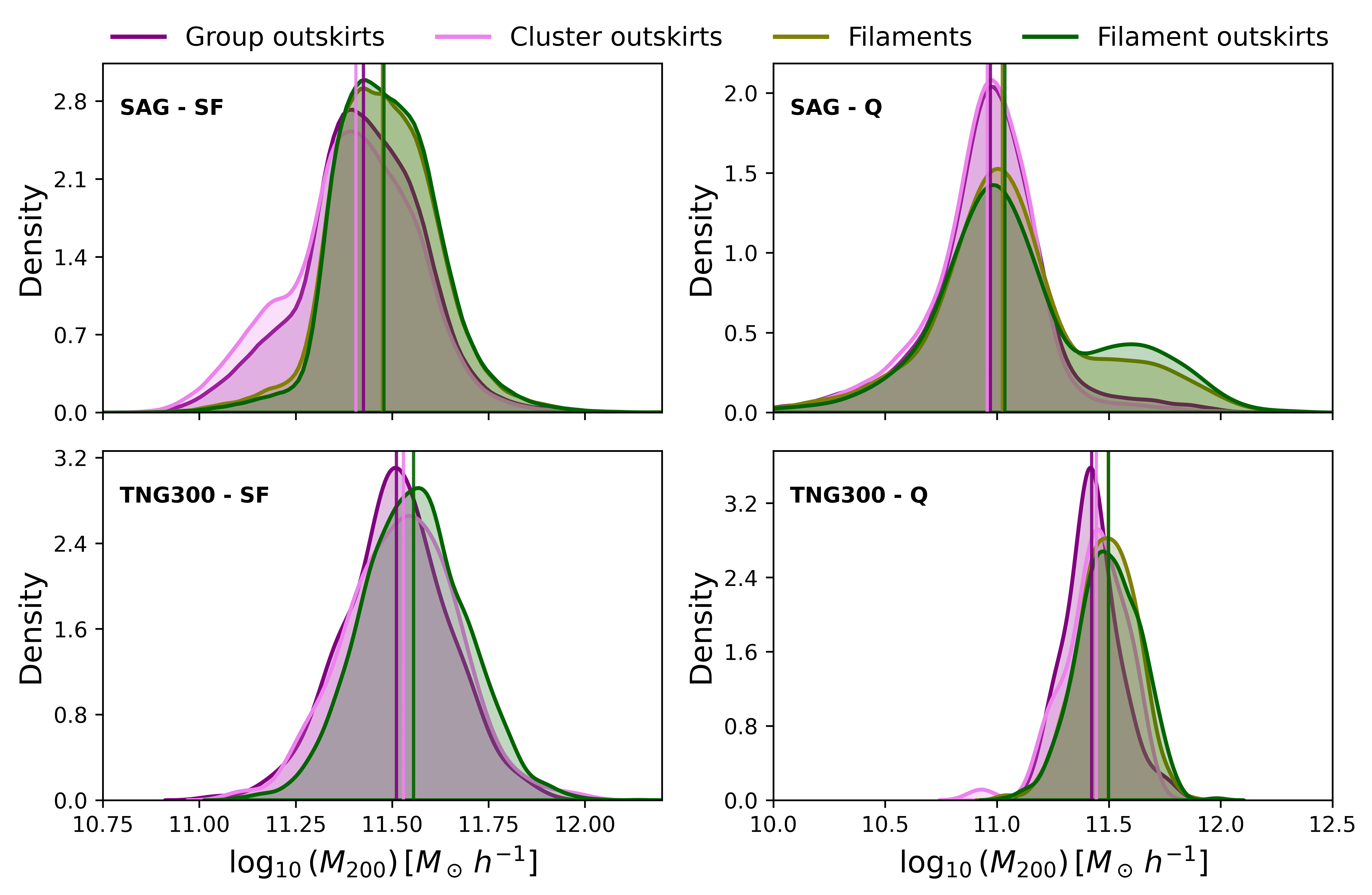}
  \caption{Probability distribution of M$_{200}$ for low-mass central galaxies across different environments. The panels and colors are designed in the same way as in Fig. \ref{fig:mgas_in_diff_environments}.}
  \label{fig:m200_in_diff_environments}
\end{figure}

Figures \ref{fig:mgas_in_diff_environments}, \ref{fig:ssfr_in_diff_environments} and \ref{fig:m200_in_diff_environments} show the distributions of gas mass, sSFR, and dark matter content, respectively, for galaxies located in group outskirts, cluster outskirts, filaments, and filament outskirts. Each panel presents the probability distribution of the corresponding galaxy property, separated by star formation activity and model. Star-forming and quenched galaxies are shown in the left and right panels, respectively. The upper and lower panels correspond to the SAG and TNG300 models, respectively, as indicated in each figure. For consistency, we maintain the same color scheme given in Fig. \ref{fig:confin_diff_environments} for each environment. For each population and environment, the median value of the corresponding property is indicated.

Star-forming galaxies in both models exhibit similar gas mass distributions across all environments, with median values around $\log_{10}(M_{\rm gas}) \sim 10.55~h^{-1}M_{\odot}$. These similarities are most evident in filamentary regions. In contrast, the distributions in group and cluster outskirts are shifted toward lower gas masses in both models, with a more pronounced separation in TNG300 relative to galaxies residing in filamentary regions. The reduced gas mass in the outskirts of massive structures is consistent with the expectation that galaxies approaching the vicinity of these systems may experience a gradual reduction in their gas supply, whereas galaxies in filaments tend to exhibit larger gas reservoirs and sustain ongoing star formation. Although the affected population appears to be relatively small, it could account for  the residual conformity signal produced by galaxies in these environments. 

In SAG, quenched galaxies exhibit systematically lower gas masses across all environments, with median values around $\log_{10}(M_{\rm gas}) \sim 9.9 - 10~h^{-1}M_{\odot}$. At fixed stellar mass and after excluding post-processed galaxies, these results indicate that quenched galaxies retain substantially smaller gas reservoirs than their star-forming counterparts, regardless of environment.
In contrast, the quenched population in TNG300 exhibits higher median gas masses, $\log_{10}(M_{\rm gas}) \sim 10.1-10.4~h^{-1}M_{\odot}$, corresponding to an offset of approximately $0.2-0.4$ dex relative to SAG. This difference indicates that quenched galaxies in the semi-analytic model retain substantially less gas than their counterparts in the hydrodynamical simulation, consistent with the stronger gas depletion histories reported by \cite{Palma_2025}, suggesting that the remaining gas reservoirs are consumed without significant replenishment.
Furthermore, in TNG300, quenched galaxies located in the outskirts of massive structures exhibit lower gas contents than those residing in filamentary regions. This environmental separation suggests that gas depletion may be more efficient in the vicinity of massive systems than within filaments. 

The sSFR distributions (Fig. \ref{fig:ssfr_in_diff_environments}) support the trends observed for the gas content. Star-forming galaxies located in group and cluster outskirts show distributions skewed toward lower sSFR values in SAG, and broader distributions in TNG300. The tail toward lower sSFR values within the star-forming population suggests the presence of transition galaxies toward quiescence. Their reduced star-formation activity could be associated with gradual gas depletion processes that can consume the available gas reservoir over long timescales without efficient replenishment (\citealt{Peng_2015}), as galaxies approach groups and clusters. Since former satellites have been excluded from our sample, this trend is unlikely to be associated with environmental processes during a previous satellite phase. Instead, it may reflect the influence of the large-scale environment surrounding massive structures. The effect appears slightly stronger in the cluster outskirts than in the group outskirts, potentially indicating a dependence on the mass of the nearby structure.

Overall, quenched galaxies are more strongly suppressed in SAG than in TNG300 (see right panels of Fig. \ref{fig:ssfr_in_diff_environments}). The median sSFR values differ by approximately $0.1 - 0.2$ dex between the two models, with the smallest difference ($\sim$0.08 dex) observed in cluster outskirts. While TNG300 shows relatively Gaussian distributions, SAG exhibits distributions skewed toward lower sSFR values across all environments. This behavior is consistent with the lower gas contents measured for quenched galaxies in the semi-analytic model.

The distribution of dark matter content (Fig. \ref{fig:m200_in_diff_environments}) reveals further environmental trends. For star-forming galaxies, both models show similar halo mass distributions across environments, with median values around $\log_{10}(M_{200}) \sim 11.4 - 11.6~h^{-1}M_{\odot}$, with slightly lower median $M_{200}$ values in group and cluster outskirts than in filamentary regions. This trend is more evident in SAG, where the reduced median content of dark matter for these galaxies suggest that, galaxies (especially in the low-mass regime) approaching to groups and clusters are more susceptible to the gravitational field of these structures, where gravitational perturbations could affect their dark matter halos (e.g., \citealt{Van_den_Bosch_2018}; \citealt{Choque-Challapa_2019}, thereby reducing their content. This hypothesis is consistent with the skewed distribution toward lower values observed for these environments in SAG. 

For quenched galaxies, the differences in the dark matter content between models are more pronounced. In TNG300, quenched galaxies closely follow the halo mass distribution of star-forming galaxies, with median values around $\log_{10}(M_{200}) \sim 11.4 - 11.5~h^{-1}M_{\odot}$. In SAG, however, quenched galaxies systematically inhabit lower-mass halos at fixed stellar mass, with median values reaching $\log_{10}(M_{200}) \sim 11 ~h^{-1}M_{\odot}$ in group outskirts. This result is particularly interesting given that quenched galaxies in SAG are preferentially located closer to massive structures than their star-forming counterparts (\citealt{Palma_2025}).

We further explore whether the systematically lower M$_{200}$ values in SAG relative to TNG300 could be associated with the abundance and spatial distribution of massive halos. 
To quantify the role of tidal interactions produced by massive interacting neighbors, we measure the contribution of merger candidate systems, namely interacting group-group, cluster-cluster, or group-cluster systems. These systems satisfy the condition $d \leq r_{1} + r_{2}$ where $d$ is the separation between the two massive systems, and $r_{1}$ and $r_{2}$ their virial radii. We examine galaxies located at distances $\lesssim 0.1~h^{-1}$ Mpc (to amplify a possible effect) and between $2$ and $3~h^{-1}$ Mpc from these systems. We evaluate both the conformity signal and the median $M_{200}$ values. The analysis presented in Appendix \ref{sec:appendixG} shows that restricting the sample to galaxies within $\lesssim0.1~h^{-1}$Mpc of interacting massive systems enhances the conformity amplitude up to $\lesssim4~h^{-1}$Mpc. However, the median $M_{200}$ values remain unchanged. This suggests that tidal forces amplify the signal without directly reducing the halo mass. We discuss the implications of these results in the following section.

In filamentary regions, the halo mass distributions of quenched galaxies in SAG exhibit an apparent bimodal distribution. One component is centered at halo masses typical of the quenched galaxy population in TNG300 ($\log_{10}(M_{200}) \sim 11.4 - 11.5~h^{-1}M_{\odot}$), while a second, more populated component appears at lower masses ($\log_{10}(M_{200}) \sim 11~h^{-1}M_{\odot}$). The presence of this low-$M_{200}$ population is particularly notable given that former satellites have been excluded from the analysis and that the sample spans a narrow stellar mass range. This finding reveals the existence of distinct populations of low-mass central quenched galaxies in filamentary environments in the model. This suggests that filaments may play a role in this population of galaxies. For instance, one possible interpretation is that the lower-mass halo population is linked to differences in halo assembly histories and to the influence of the filamentary environment on halo growth \citep{Borzyszkowski_2017}. We discuss this possibility in more detail in the next section.


\section{Discussion and Conclusions} \label{sec:discussion}

In this work, we examine the large-scale conformity signal produced by a population of low-mass central galaxies at $z=0$ after excluding post-processed galaxies. We study this phenomenon in two galaxy formation models, TNG300 and SAG. Although part of the signal can be explained by backsplash objects, a residual conformity signal persists \citep{Ayromlou_2023, Palma_2025}. Here, we investigate the origin of this residual signal and its connection to the large-scale environment.

The different impact of former satellites on the conformity signal, revealed after removing this population (see the black lines in Fig. \ref{fig:confin_diff_environments}) is consistent with the different fractions of former satellite galaxies identified in each model ($\sim 17\%$ in SAG against $\sim 45\%$ in TNG300). This result highlights the regime in which differences in numerical resolution and galaxy formation prescriptions become more evident. However, these differences disappear when the conformity signal is measured using the full population of low-mass central galaxies, for which both simulations exhibit a quantitatively comparable signal (\citealt{Palma_2025}). These results suggest that the conformity signal is broadly robust against differences in the galaxy formation prescriptions adopted by the two models. The persistence of a residual signal in both models indicates that the physical explanation of the correlation at large-scales cannot be exclusively attributed to pre-processing effects.

\citet{Lacerna_2022} found that galaxies inhabiting the vicinity of galaxy groups and clusters, defined in terms of megaparsec scales from their centers, are the primary drivers of the large-scale conformity signal. These regions are well-known to host backsplash objects, consistent with their known role in driving the signal. The contribution of these galaxies has been found to be more prominent in hydrodynamical models, such as TNG300, than in semi-analytic models (e.g., L-galaxies in \citealt{Ayromlou_2023}, and SAG in \citealt{Palma_2025}). 

We found that the residual signal can be robustly traced by the cosmic web. Galaxies in the outskirts of groups and clusters contribute at small separations, but their impact declines rapidly beyond a few megaparsecs ($\sim 1.8~h^{-1}$Mpc). In contrast, filamentary regions sustain the signal over a broader range of separations (up to $\sim 7~h^{-1}$Mpc in SAG and $\sim 4~h^{-1}$Mpc in TNG300). Therefore, filaments emerge as key regions that trace the large-scale conformity signal, whose effect is not only confined to the filament spine, but also in the immediate surrounding environments of such structures, which show to closely reproduce the shape and amplitude of the residual signal, despite the small fraction of galaxies in these environments (see Table \ref{tab:gals_TNG_SAG_in_diff_environments}). 

For TNG300, given the non-negligible uncertainties associated with the smaller sample, we tested the conformity signal in filamentary regions by including former satellite galaxies to increase the statistics (see Appendix \ref{sec:appendixC}), and taking advantage of the role we know these objects contribute to the signal. The resulting signal closely resembles that measured for the full population of low-mass central galaxies when no environmental classification is applied, providing additional support for the interpretation that filamentary regions contribute significantly to the large-scale conformity signal. The persistence of a qualitatively similar, although weaker signal after removing former satellites further indicates that galaxies residing in filaments contribute to the conformity signal, independently of pre-processing effects, as shown in Fig. \ref{fig:confin_diff_environments}. Furthermore, the reduction in amplitude observed after removing former satellites is consistent with filamentary regions hosting a non-negligible population of such objects\footnote{Studies such as \cite{Salerno_2022}, \cite{Navdha_2025} have found that there is a population of former satellites identified in filamentary structures, where the accretion of galaxies through filaments could occur.}. However, the analysis of the fractions of former satellites per environment is beyond the scope of this work.

This finding supports the scenario in which the cosmic web mediates environmental effects well beyond the virial radius of massive halos, as they connect the virialized structures with the surrounding LSS. 

Although the identification of filamentary structures depends on the adopted tracer population and the persistence threshold used in DisPerSE, the conformity signal associated with filaments remains qualitatively consistent in both simulations. The two models employ different configurations in the extraction of the filamentary network (see details in Sec. \ref{sec:method}), which could partly account for the variations in the fraction of galaxies identified within filaments (see left panel of Fig. \ref{fig:hist_gals_in_diff_environments}). Despite these differences, the median linear densities of the two filament populations are quite similar between the models (see Tab. \ref{tab:properties_filament_population}), and filamentary regions trace a comparable residual conformity signal in both cases. 

We further tested the robustness of this result by modifying the adopted filament thickness and by evaluating the impact of galaxies located in overlapping filamentary regions. Reducing the filament thickness from $1~h^{-1}$Mpc to $0.5~h^{-1}$Mpc shows that the signal remains robust under changes in this parameter. Similarly, galaxies in overlapping filamentary regions do not drive the observed signal, as the conformity amplitude remains consistent regardless of whether a single or multiple filaments are associated with a given galaxy. These tests demonstrate that the connection between filamentary environments and the residual conformity signal is not driven by a particular choice of filament definition. 

Beyond the robustness of the filament identification, the conformity signal exhibits a clear dependence on filament properties. The signal remains detectable in both long and short filaments (see Fig. \ref{fig:conf_SAG_short_vs_large_fil}), as well as in low-linear and high-linear density filaments (see \ref{fig:conf_SAG_high_vs_low_density_fil}), indicating that the correlation is a generic feature of filamentary environments rather than being restricted to a particular filament population. However, systematic differences in amplitude are observed. Dense linear filaments exhibit stronger conformity than low-linear density filaments, while short filaments show enhanced amplitudes relative to long filaments. Since short filaments are known to be systematically denser than long filaments \citep{DGE_2020}, both trends point to a common interpretation in which the local density of the filamentary environment modulates the strength of the conformity signal.

A similar trend is found when filaments are separated according to their mass. High-mass filaments ($\geq~10^{13.39}~h^{-1}M_{\odot}$) exhibit stronger conformity compared to low-mass filaments (Appendix \ref{sec:appendixF}). These results further support a scenario in which denser filaments are more effective at inducing the correlations in star-formation activity observed over megaparsec scales.

A plausible explanation for the enhanced conformity signal in dense and short filaments is that environmental processes that suppress gas accretion or enhance tidal interactions become more efficient in these regions. For instance, denser filaments may be more effective at inducing cosmic web stripping \citep{Benavides_2025}, although this mechanism's efficiency also depends on the relative velocity of galaxies. We further note that the SAG filament network, traced using luminous galaxies ($M_{\rm r} < -21$), preferentially recovers intrinsically denser and more luminous filaments (see Table \ref{tab:properties_filament_population}). This tracer choice likely narrows the dynamic range between our high- and low- linear density subsamples, and could explain why the difference in conformity amplitude between these subsamples remains moderate in the model.

The origin of the weak conformity signal detected in low-density regions in SAG (upper panel Fig. \ref{fig:confin_OE}), remains unclear and is beyond the scope of this paper to analyze in detail. Nevertheless, several effects may contribute to this residual signal. Although the DisPerSE algorithm provides robust identification of filamentary structures, very low-density regions, such as voids, may be subject to some degree of misclassification. To evaluate the possible influence of massive neighbors on the signal found in these environments, we identified the fraction of massive companions of galaxies classified in other environments within regions of $0.8$ and $0.5~h^{-1}$Mpc. We found significantly lower fractions of massive companions at these distances, ranging from $6-9\%$ for quenched galaxies and from $1-2\%$ for star-forming galaxies, indicating a well defined, though not entirely clean, selection of low-density regions by the algorithm. As a follow-up work, we will investigate conformity in void environments using dedicated void finder algorithms (e.g., Sparkling \citealt{Ruiz_2026}; REVOLVER \citealt{Seshadri_2019}; popcorn \citealt{Dante_2023}).

To search for possible physical mechanisms behind the conformity signal, we analyzed the distributions of gas mass, sSFR, and dark matter content for galaxies in each large-scale environment (see Figs. \ref{fig:mgas_in_diff_environments}, \ref{fig:ssfr_in_diff_environments}, \ref{fig:m200_in_diff_environments}). In particular, we found that star-forming galaxies in the outskirts of massive structures show distributions skewed toward lower sSFR values and, consequently, lower gas mass content. This behavior could indicate slow quenching processes as galaxies move toward the clusters. No significant reduction in gas mass is found in filamentary regions; on the contrary, galaxies in filaments seems to retain larger gas reservoirs, consistent with the scenario in which filaments act as channels for cold gas supply that sustain ongoing star formation \citep{DGE_2021, Salerno_2022}.
The only difference appears in quenched galaxies in SAG, where a tail toward lower sSFR values is observed, a feature also present for galaxies in the outskirts of massive structures. We also verified the median gas content and sSFR for galaxies in different filament populations, and found these quantities remain remarkably similar across the filaments. However, galaxies in dense and short filaments exhibited a slightly more extended tail toward lower gas masses and lower sSFR values, which may suggest that galaxies in these type of filaments are more susceptible to processes that gradually reduce their outer gas reservoirs, such as cosmic web stripping \citep{Benavides_2025} or strangulation \citep{Peng_2015}.

The reduced $M_{200}$ values observed for quenched galaxies in SAG could in principle reflect a contamination from former satellite galaxies, which typically exhibit a reduced dark matter content compared to non-preprocessed galaxies. We tested this possibility by analyzing the phase-space diagrams in the model, and found no evidence of such contamination. Instead, we identify a large population of galaxies located in the first infall region, which are expected to fall into groups or clusters at later times and may also contribute to the skewed sSFR distribution observed in the model.

The bimodal $M_{200}$ distribution observed for quenched galaxies in filamentary regions in SAG suggests the coexistence of two distinct populations. One possibility is that the component at lower halo masses ($\log_{10}(M_{200}) \sim 11~h^{-1}M_{\odot}$) corresponds to early formed systems whose dark matter accretion was suppressed by the anisotropic tidal field of the embedding filament. \citet{Borzyszkowski_2017} showed that halos embedded in high-density filaments experience a strongly sheared velocity field that redirects infalling material along tangential rather than radial trajectories, effectively stalling their mass growth. This mechanism operates preferentially for low-mass halos and is directly connected to assembly bias, where early-forming halos embedded in dense filamentary structures cease to accrete dark matter, resulting in lower $M_{200}$ values at fixed stellar mass. This interpretation is consistent with the direct link between conformity and assembly bias recently identified in SAG by \citet{Lacerna_2025}.

The enhancement of conformity produced by galaxies with a nearby merger-candidate system (within $0.1h^{-1}$Mpc) supports the role of tidal fields in enhancing the signal (see Appendix \ref{sec:appendixG}), extending its influence up to $\sim 4h^{-1}$Mpc. However, the similarity in the halo mass distributions of these populations indicates that tidal interactions alone do not explain the systematically lower $M_{200}$ values of quenched central galaxies across environments in SAG. Therefore, this offset likely reflects differences in halo assembly histories. \citet{Lacerna_2025} show a direct link between conformity and assembly-type bias in SAG, indicating that central galaxies with stronger conformity also exhibit stronger assembly bias at fixed halo mass ($10^{11.6} \leq \rm M_{h}/h^{-1} M_{\odot} \leq 10^{11.8}$). In this context, early-forming halos may host galaxies that quenched more efficiently and experience reduced late-time mass accretion, leading to quenched central galaxies residing in lower-mass halos at fixed stellar mass.

This scenario is consistent with halo-galaxy co-evolution frameworks. Empirical models such as Stellar Halo Accretion Rate Coevolution (SHARC, \citealt{Rodriguez_Puebla_2016}) link galaxy growth to halo mass accretion histories, predicting assembly-dependent segregation in the stellar-to-halo mass relation (\citealt{Rodriguez_Puebla_2015}, \citealt{Rodriguez_Puebla_2016}). Although observational studies generally find that quenched centrals reside in more massive halos at fixed stellar mass (e.g., \citealt{More_2011}, \citealt{Tinker_2013}, \citealt{Rodriguez_Puebla_2015}, \citealt{Mandelbaum_2016}, \citealt{Lange_2019}), semi-analytic implementations can produce the opposite trend if early halo formation leads to efficient quenching and suppressed subsequent growth.

Our results suggest that filamentary environments could directly affect the star-formation activity of low-mass galaxies, inducing a correlation with neighboring galaxies, the referred large-scale conformity signal studied here. By analizing the gas mass distribution of galaxies in dense and short filaments, we identify a tail toward lower values, consistent with a scenario in which galaxies in these structures could be under the influence of cosmic web stripping, a phenomenon previously reported in dwarf galaxies in TNG50 \citep{Benavides_2025}. Such a mechanism may also operate, at least partially, in the galaxy populations studied here, and could contribute to the enhanced signal measured in dense and short filaments.

An observational study using galaxies from SDSS (DR18, \citealt{Almeida_2023}) with filamentary structures identified using DisPerSE (see \citealt{Rodriguez_AMD_2025}) will be addressed in future work to investigate the observed correlation in these structures. In addition, using data from SPLUS \citep{Mendes_de_oliviera_2019} and future surveys such as CHANCES \citep{chances_2025} will provide wide area coverage around galaxy clusters, enabling studies of the connection between the conformity signal and the LSS. The role of backsplash galaxies may also be explored using codes such as ROGER \citep{de_los_Rios_2021}, allowing us to evaluate their contribution to the signal in observational data.

In summary, our results demonstrate that the residual large-scale conformity signal, persisting after removing the former satellite population, is primarily traced by filamentary structures, which sustain the correlation over separations up to $\sim 7~h^{-1}$Mpc. This connection is robust against the choice of filament identification strategy, filament thickness, and the presence of overlapping structures. The strength of the signal is modulated by the density of the filament, with denser and more massive filaments producing stronger conformity. The distributions of gas mass, sSFR, and dark matter content across environments provide physical evidence for possible environmental processes acting on low-mass central galaxies independently of pre-processing, which could include gradual gas depletion and suppressed dark matter accretion in filament environments. Moreover, the systematically lower halo masses of quenched galaxies in SAG, together with the link between conformity and assembly bias \citep{Lacerna_2025}, suggest that filamentary regions may preferentially host early-forming, assembly-biased systems whose halo growth was stalled by the cosmic web. Together, these results establish filamentary structures as a key element in the environmental interpretation of the large-scale conformity signal.

\begin{acknowledgements}
The authors thank the referee for constructive comments and suggestions that helped improve the manuscript. DP acknowledges the financial support from Fundaçao de Amparo à Pesquisa do Estado de São Paulo (FAPESP; project 2011/51680-6). DP also acknowledges Andrés N. Ruiz for providing the MDPL2-SAG dataset, and useful discussions with halo-galaxy connection group, Amanda Lopes, Gissel Montaguth, and Micheli Trindade-Moura. IL acknowledges support from the ANID FONDECYT Regular grant 1261197. ADMD and CASS acknowledge support from the Universidad Técnica Federico Santa María through the Proyecto Interno Regular \texttt{PI\_LIR\_25\_04}. LSJ acknowledges the support from CNPq (308994/2021-3)  and FAPESP (2011/51680-6). MCA acknowledges financial support from ANID BASAL project FB210003. NCC acknowledges the support from FONDECYT Postdoctorado 2024, project number 3240528. FR thanks the support by Agencia Nacional de Promoción Científica y Tecnológica, the Consejo Nacional de Investigaciones Científicas y Técnicas (CONICET, Argentina) and the Secretaría de Ciencia y Tecnología de la Universidad Nacional de Córdoba (SeCyT-UNC, Argentina).
\end{acknowledgements}

\bibliographystyle{aa}
\bibliography{mibiblio}

\begin{appendix}
    
\section{Conformity signal in Other Environments
} \label{sec:appendixA}
In DisPerSE, galaxies that are not classified in the outskirts of clusters or groups, or filamentary regions, belong to other environment, where voids and walls are included. Given that our interest along the work was to study filamentary regions, and we found that the density plays an important role in shaping the signal, as a complement test we study galaxies in other environments, as an approach to galaxies in low-density regions. A deeper study in voids would be addressed in future work.

\begin{figure}[htbp]
  \centering
    \includegraphics[width=0.5\textwidth]{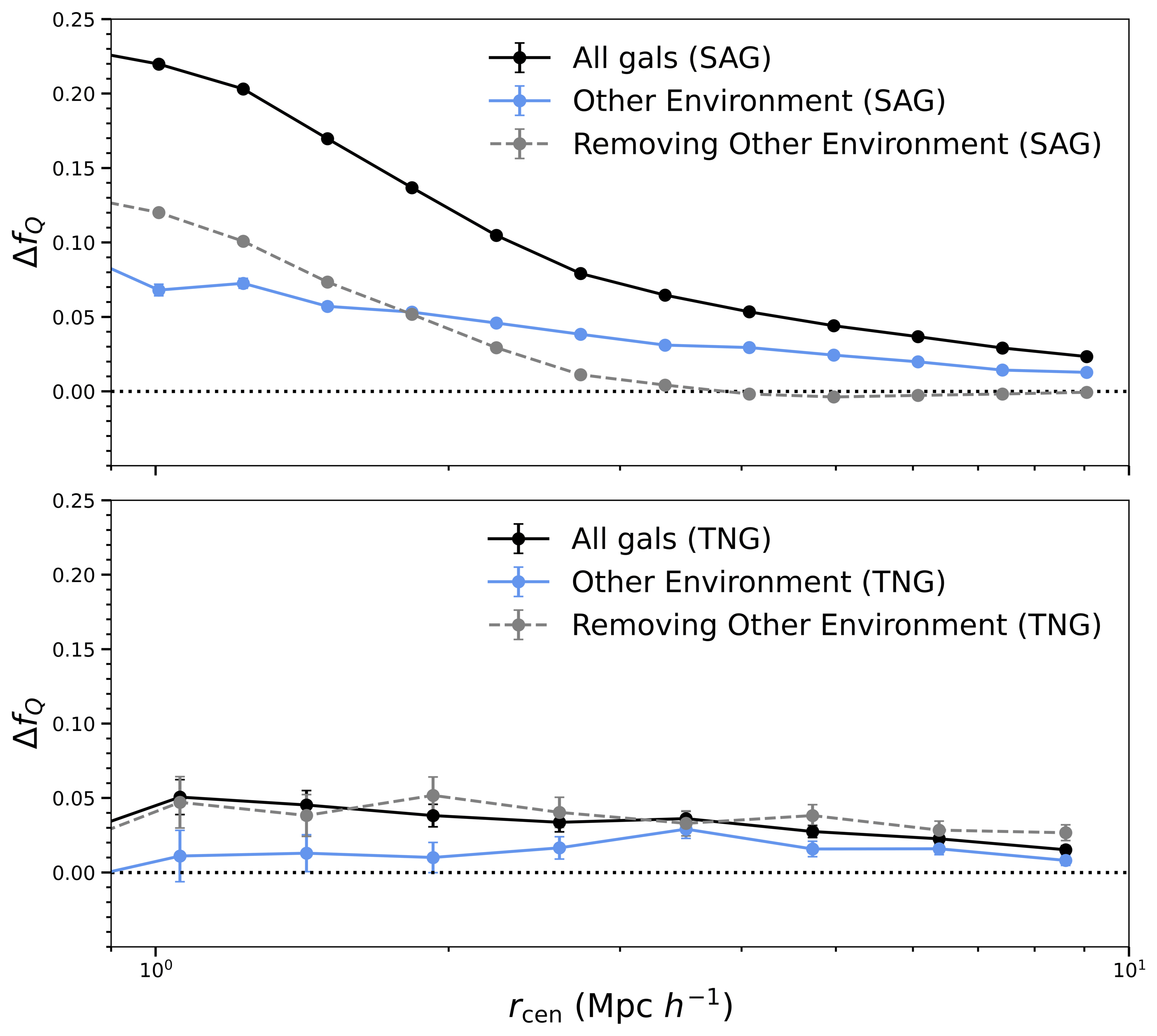}
  \caption{Conformity signal, $\Delta f_{\rm Q}$, for galaxies in other environments (light blue solid line) for SAG (upper panel) and TNG300 (lower panel). The signal measured in these environments implicitly correspond to the sample where galaxies in the outskirts of groups and clusters, as well as, filamentary regions are excluded. In SAG, a weaker signal still persist, whereas no significant contribution is observed in TNG300. In addition, we trace the measurements of the signal without galaxies in other environments (gray dashed line) for each model.}
  \label{fig:confin_OE}
\end{figure}

Figure \ref{fig:confin_OE} shows the conformity signal produced by galaxies in other environments (light blue solid line), for SAG (upper panel) and TNG300 (lower panel). Galaxies in other environments constitute the largest populations in the models (see Fig. \ref{fig:hist_gals_in_diff_environments}), which explains their smaller uncertainties.

In both cases, galaxies in other environments exhibit a weaker signal, that translates to a significantly lower amplitude across all separations. In SAG, even when the amplitude is substantially reduced with respect to the signal measured in the other large-scale environments, it remains different from zero over the full separations analyzed. In addition, the light blue line implicitly corresponds to the sample of low-mass central galaxies without galaxies in the outskirts of massive structures, or filamentary regions, reinforcing the contribution of these environments to the measured signal. In TNG300, the conformity amplitude is likewise generally weaker than the global signal, particularly at separations below $\sim 2~h^{-1}$Mpc. At intermediate separations ($\sim 3 - 4~h^{-1}$Mpc, however, the signal partially recovers and approaches the amplitude measured for the full galaxy sample before declining again at large distances. The corresponding $f_{\rm Q}$ measurements support this analysis, showing modest differences.

\section{Quenched fractions around low-mass central galaxies
} \label{sec:appendixB}

\begin{figure*}[h!]
  \centering
    \includegraphics[width=\textwidth]{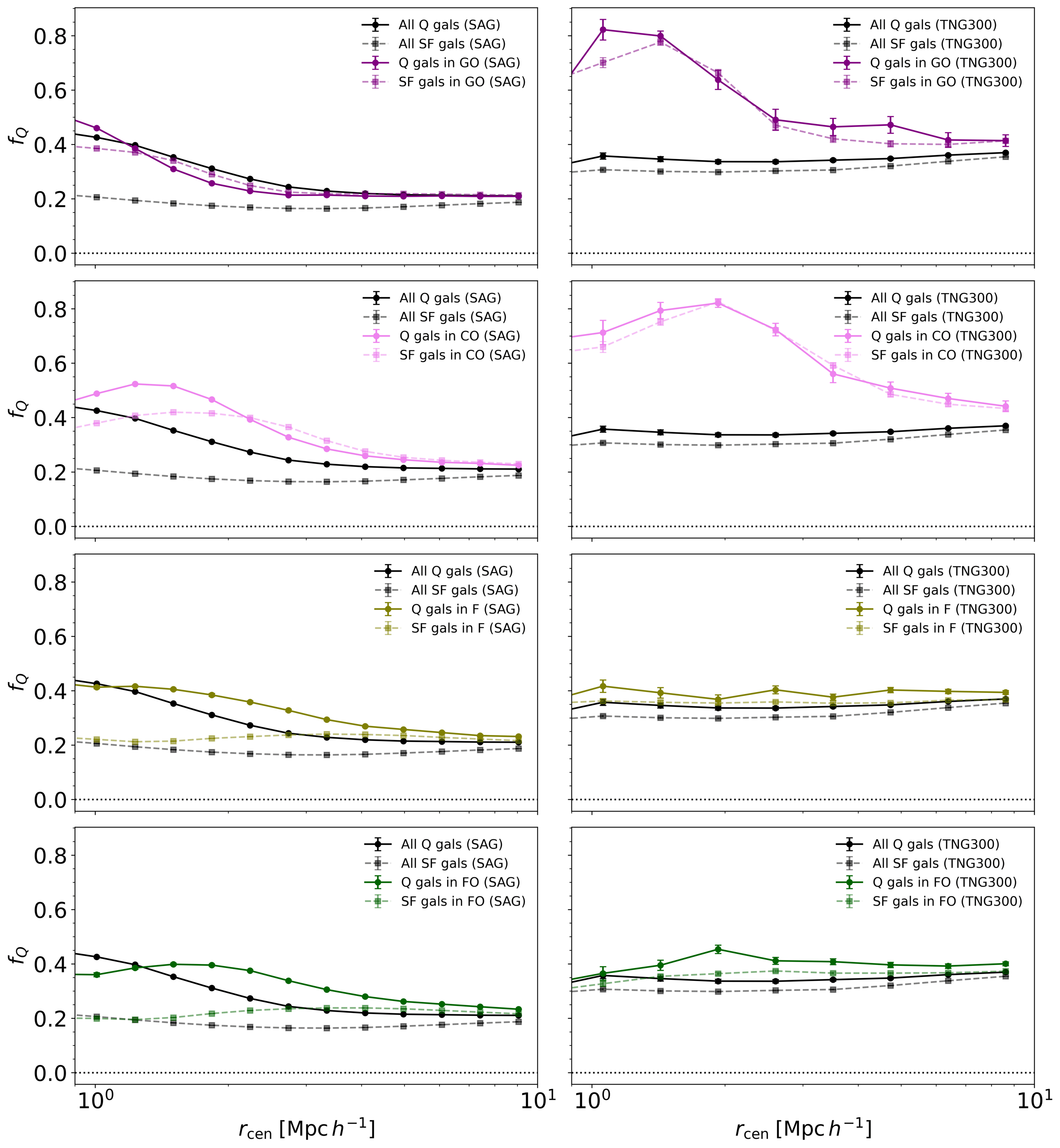}
  \caption{The mean quenched fraction, $f_{\rm Q}$, of neighboring galaxies is shown for both quenched (black solid line) and star-forming (gray dashed line) low-mass central galaxies. The colored lines represent the $f_{\rm Q}$ of quenched (solid lines) and star-forming (dashed lines) central galaxies separated in group outskirts (GO), cluster outskirts (CO), filaments (F), and filament outskirts (FO). The figure is colored and organized in the same way as Fig. \ref{fig:confin_diff_environments}.}
  \label{fig:fQ_in_diff_environments}
\end{figure*}

Figure \ref{fig:fQ_in_diff_environments} shows the quenched fraction, $f_{\rm Q}$, of neighboring galaxies around low-mass central galaxies out to 10 $h^{-1}$Mpc. The black solid and gray dashed lines represent the $f_{\rm Q}$ around quenched and star-forming central galaxies, respectively, independent of their environment. The left and right panels correspond to SAG and TNG300 models, respectively. We use the same color designation shown in Fig. \ref{fig:confin_diff_environments} for each environment, and solid (with filled circles) and dashed (with squares) lines to distinguish the $f_{\rm Q}$ around quenched and star-forming central galaxies, respectively.

\FloatBarrier

Looking at the left panels of the figure (SAG model), we find a clear excess in the quenched fraction around quenched central galaxies located in cluster outskirts (CO), particularly in the first distance bins. This excess directly translates into the conformity signal observed in this environment in Fig. \ref{fig:confin_diff_environments}. However, the largest differences in the quenched fraction around quenched and star-forming central galaxies, responsible for the conformity signal, are found in filamentary regions, that is, filaments (F) and filament outskirts (FO), where the signal remains detectable out to larger distances.
This behavior is consistent with the strong conformity signal associated with this population.

The right panels of Fig. \ref{fig:fQ_in_diff_environments} show that, for galaxies in group outskirts, a detectable signal within the first distance bin, that is, around $\sim 1~h^{-1}$Mpc can be detected, aligned with the results obtained in Fig. \ref{fig:confin_diff_environments}. In cluster outskirts instead, the quenched fraction exhibits a very similar mean values across distance bins around quenched and star-forming central galaxies. This indicates that the contribution of this population to the conformity signal in TNG300 is weak, as shown in Fig. \ref{fig:confin_diff_environments}, where the slight variations in $f_{\rm Q}$ remain within the uncertainties. 
In filamentary regions, differences in $f_Q$ between quenched and star-forming galaxies remain detectable, but they are significantly weaker than those found in SAG. As a consequence, a subtle conformity signal is observed at $\sim 3-4~h^{-1}$~Mpc, as shown in Fig. \ref{fig:confin_diff_environments}.

\section{Conformity signal in filaments including former satellite galaxies} \label{sec:appendixC}

We examine the conformity signal of low-mass central galaxies residing in filamentary regions by including former satellite galaxies in TNG300. This approach allows us to assess whether the behavior of the signal in filaments is affected by sample size limitations in the simulation.

\begin{figure}[htbp]
  \centering
    \includegraphics[width=0.5\textwidth]{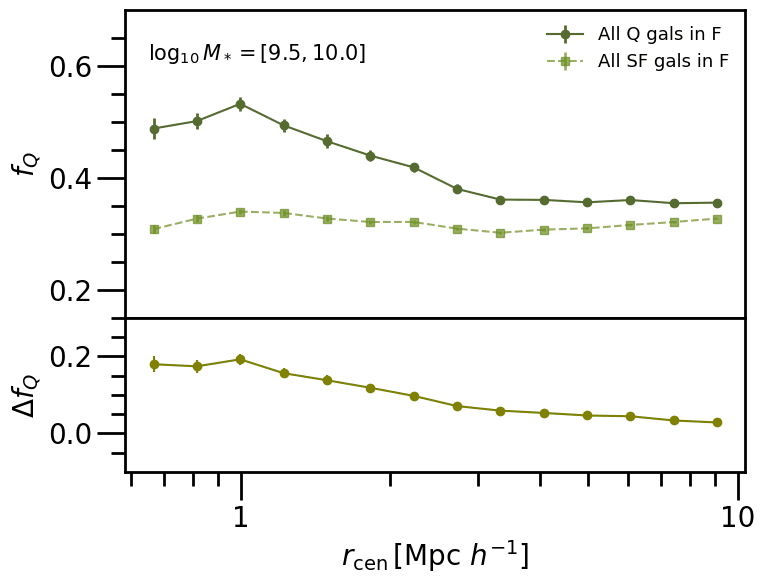}
  \caption{Quenched fraction $f_{\rm Q}$ of neighboring galaxies around all low-mass central galaxies in filamentary regions, including the former satellite population in TNG300. The lower panel shows the conformity signal, $\Delta f_{\rm Q}$. Similar mean values were previously reported by \citealt{Lacerna_2022} and \citealt{Palma_2025}, when no environmental separation was applied.}
  \label{fig:confin_fil_incl_backsplash}
\end{figure}

\FloatBarrier

The upper panel of Fig. \ref{fig:confin_fil_incl_backsplash} shows the quenched fraction around quenched and star-forming low-mass central galaxies in filaments, whereas the lower panel presents the corresponding conformity signal, $\Delta f_{\rm Q}$. The mean values observed in this figure are very similar to those reported in previous studies of the same stellar mass range in the model (e.g. \citealt{Lacerna_2022}, \citealt{Wang_2023}, \citealt{Palma_2025}). 
This result indicates that galaxies in filaments closely reproduce the total conformity signal obtained when all low-mass central galaxies are considered, independent of environment. The strength of the signal, even when former satellites are included, indicates that environments such as filaments drive the large-scale conformity signal. After removing former satellite galaxies, the signal decreases in amplitude, as shown in the black solid line in the right panels of Fig. \ref{fig:confin_diff_environments}. This test therefore supports the influence of these environments in driving the correlation of the star-formation activity of low-mass central galaxies with its neighbors, as well as implicitly exhibits that a remarkable population of former satellites in TNG300 are embedded in filaments.

A similar behavior is observed in SAG (see left panels of Fig. \ref{fig:confin_diff_environments}), where galaxies in filamentary regions closely reproduce the amplitude and shape of the total signal. In TNG300, the signal seems to exhibit a similar trend, but it remains noisier due to the smaller number of galaxies in these regions as presented in Sec. \ref{sec:results}. 
These results support the idea that galaxies in filaments contribute to shaping the large-scale conformity signal. Furthermore, this test confirms that the signal previously found for galaxies residing in filaments (right panel of Fig. \ref{fig:confin_diff_environments}) is not a statistical artifact driven by the small size of this population, but rather reflects a real contribution to the overall conformity signal.

\section{Conformity signal by removing population of galaxies} \label{sec:appendixD}

\begin{figure*}[htbp]
  \centering
    \includegraphics[width=\textwidth]{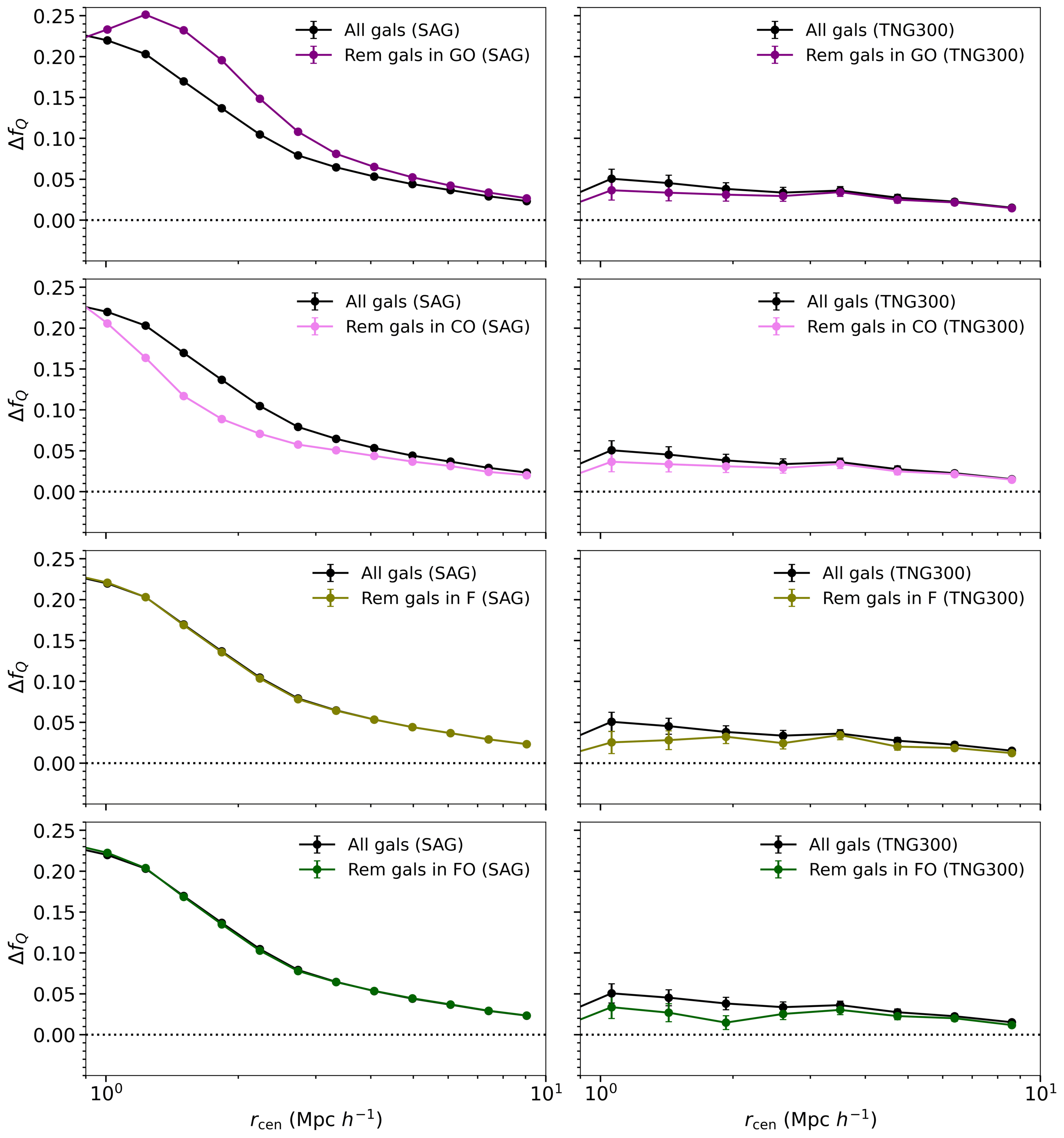}
  \caption{Conformity signal for low-mass central galaxies without former satellite objects (black solid line), compared with the signal obtained after removing the galaxies per environment (colored lines). From top to bottom: removing galaxies in group outskirts (purple), cluster outskirts (violet), filaments (olive), and filament outskirts (dark green). Left panels: semi-analytic model, SAG. Right panels: hydrodynamical model, TNG300.}
  \label{fig:confin_diff_environments_by_removing_per_pop}
\end{figure*}

Following the approaches of \cite{Lacerna_2022}, \cite{Ayromlou_2023}, \cite{Wang_2023}, and \cite{Palma_2025}, we evaluate the impact of removing galaxies by environment. That is, we excluded populations of galaxies (e.g., galaxies in group outskirts) to measure how the signal varies in each case. Fig. \ref{fig:confin_diff_environments_by_removing_per_pop} shows that removing galaxies in filamentary regions produces negligible changes in the measured signal. This behavior is more evident in SAG than in TNG300, where in the latter reduced mean values are observed, although within the uncertainties. This suggests that these galaxies do not dominate the total conformity signal in SAG. However, when we evaluate the signal separately for each galaxy population (as shown in the left panels of Fig. \ref{fig:confin_diff_environments}), we find that low-mass central galaxies in filamentary regions exhibit a clear intrinsic conformity signal in the model. The relatively small fraction of galaxies in these environments ($\sim 5\%$ vs $\sim 25\%$ in SAG and TNG300, respectively) could naturally explain why their removal does not lead to a significant modification of the total signal. 

In contrast, removing galaxies in other environments (see gray dashed lines in Fig. \ref{fig:confin_OE}) substantially modifies the signal in SAG, indicating a strong influence of this population on the measured conformity. 
This reflects the dominant statistical weight of these galaxies, which is approximately 80\% in both models, despite their weak intrinsic conformity signal (see Fig. \ref{fig:confin_diff_environments}).

\FloatBarrier

\section{Conformity signal on filament properties in TNG300} \label{sec:appendixE}
\begin{figure}[htbp]
  \centering
    \includegraphics[width=0.5\textwidth]{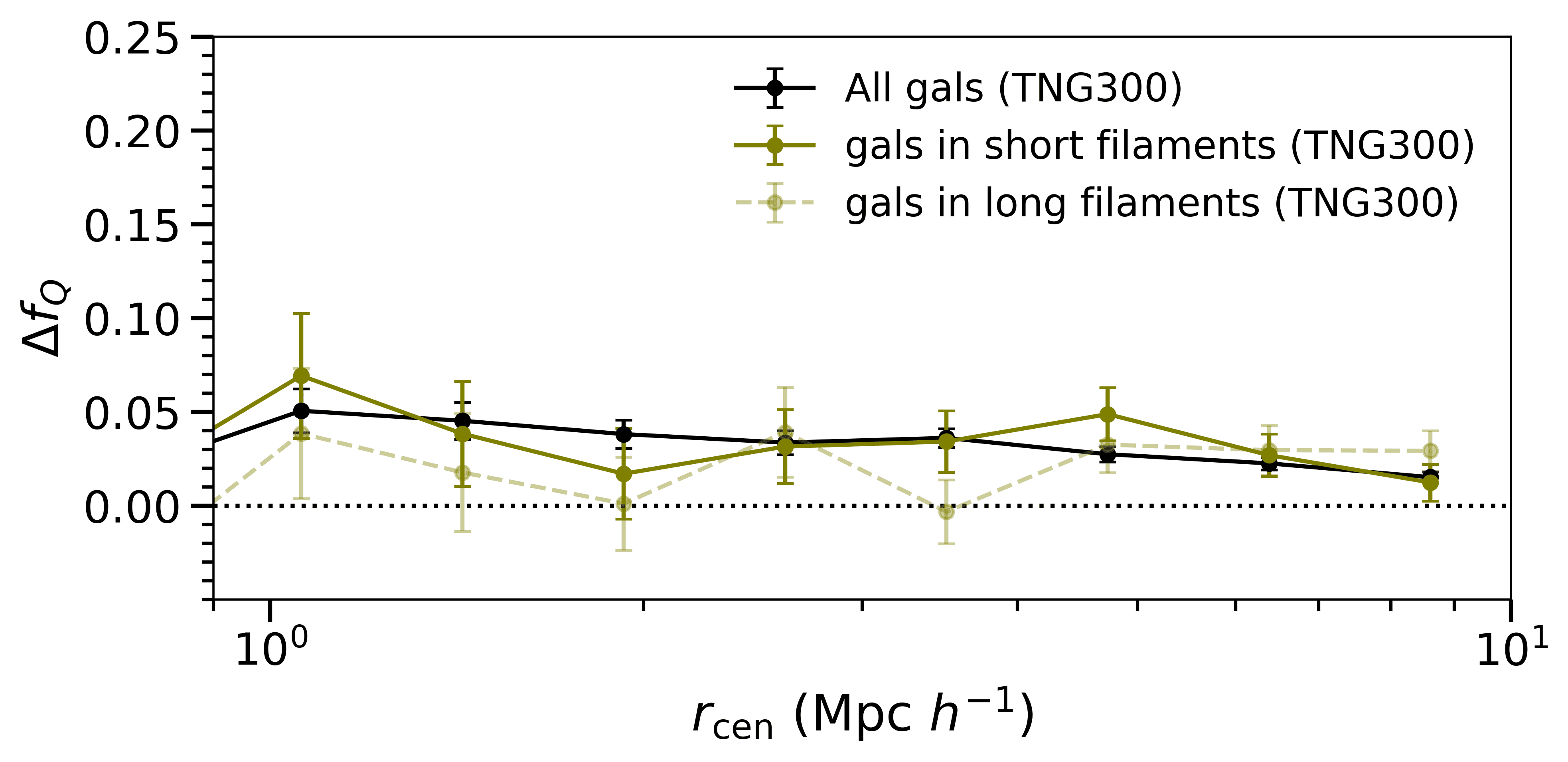}
  \caption{Conformity signal, $\Delta f_{\rm Q}$, for galaxies in long and short filaments in TNG300.}
  \label{fig:conf_TNG_short_vs_large_fil}
\end{figure}

\begin{figure}[htbp]
  \centering
    \includegraphics[width=0.5\textwidth]{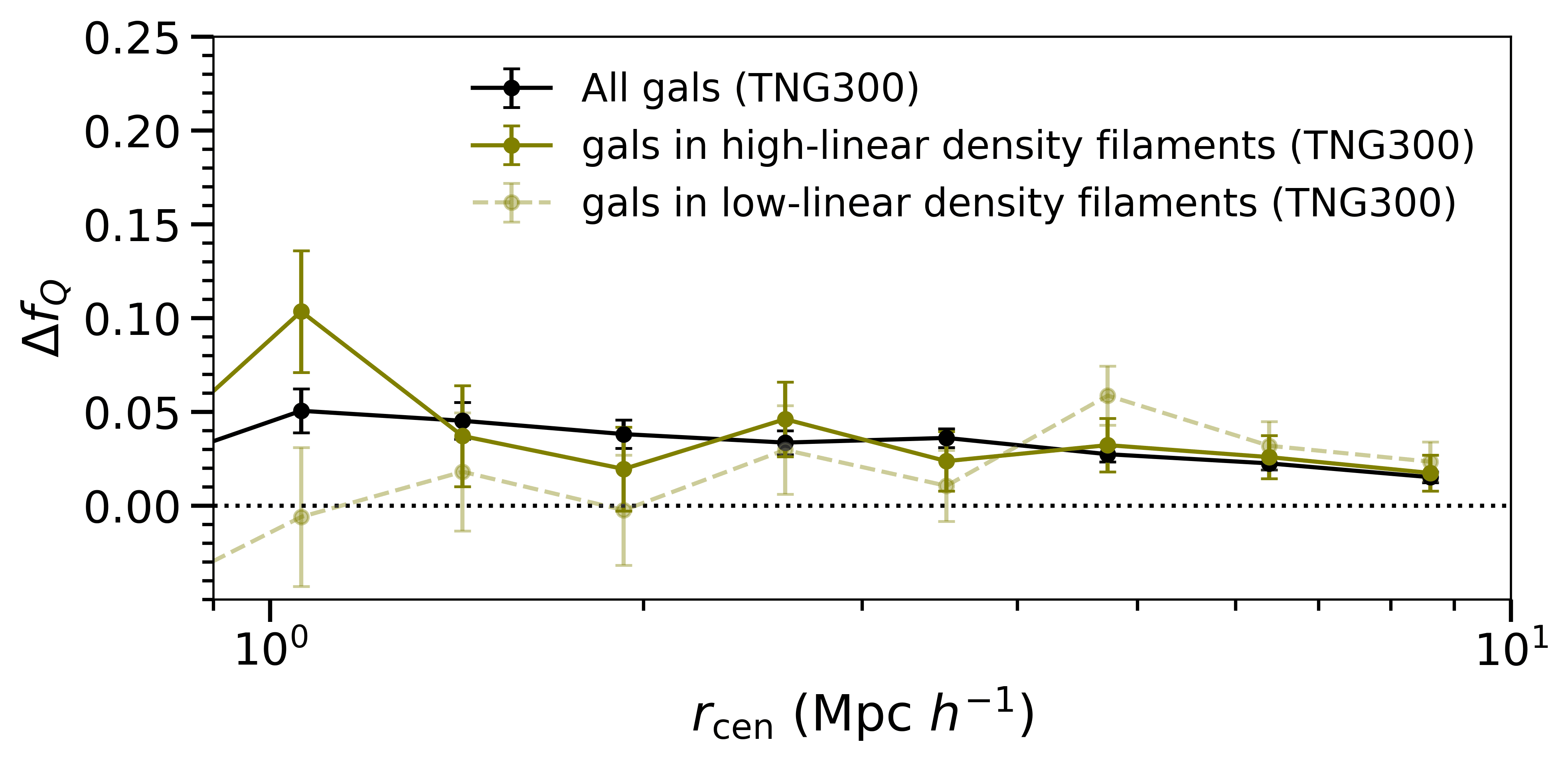}
  \caption{Conformity signal, $\Delta f_{\rm Q}$, for galaxies in high-linear density and low-linear density filaments in TNG300.}
  \label{fig:conf_TNG_high_vs_low_density_fil}
\end{figure}

We repeat the analysis performed for SAG (see Figs. \ref{fig:conf_SAG_short_vs_large_fil} and \ref{fig:conf_SAG_high_vs_low_density_fil}), evaluating the large-scale conformity signal produced by galaxies in filaments, and how the properties of these structures could contribute to the overall signal observed in Fig. \ref{fig:confin_diff_environments}. Figures \ref{fig:conf_TNG_short_vs_large_fil} and \ref{fig:conf_TNG_high_vs_low_density_fil} show the conformity signal produced by galaxies located in long and short filaments, as well as, high-density and low-density filaments, respectively, for TNG300. We split the population of filaments using the median values of the length and linear density detailed in Table \ref{tab:properties_filament_population}.

We found similar results in terms of density. That is, galaxies embedded in high-linear density filaments enhance the correlation of galaxies with its neighbors compared to galaxies in low-linear density filaments. The same occurs for galaxies inhabiting short filaments, where the signal produced by these galaxies became larger than for galaxies in long filaments. In both cases, the differences are not substantial, in part due to the small statistical sample. However, by looking the mean values, it is possible to see a larger contribution produced by dense filaments, in good agreement with the results obtained in the semi-analytic model.

\section{Dependence of the conformity signal on filament mass} \label{sec:appendixF}

As a follow-up to the analysis presented in Appendix \ref{sec:appendixC}, and Appendix \ref{sec:appendixE}, we investigate the conformity signal in filamentary regions and its dependence with the mass of the hosting filament, considering all low-mass central galaxies in TNG300, for a statistical robustness analysis. We separate galaxies according to whether they reside in low- or high-mass filaments. The mass of the filament is estimated by counting the dark matter particles located within a cylinder of radius $2~h^{-1}$Mpc around the filament axis and multiplying this number by the particle mass. The median mass of the filaments is $M_F= 10^{13.39}~h^{-1}M_{\odot}$, which was adopted as a threshold to define low-mass filaments (with masses below the median) and high-mass filaments (with masses above the median).

\begin{figure*}[htbp]
  \centering
  \begin{subfigure}[b]{0.49\textwidth}
    \centering
    \includegraphics[width=\textwidth]{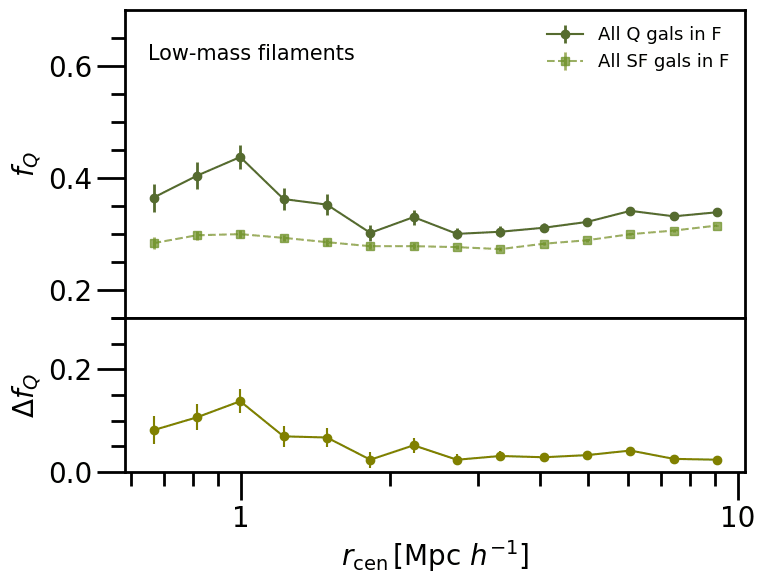}
  \end{subfigure}
  \begin{subfigure}[b]{0.49\textwidth}
    \centering
    \includegraphics[width=\textwidth]{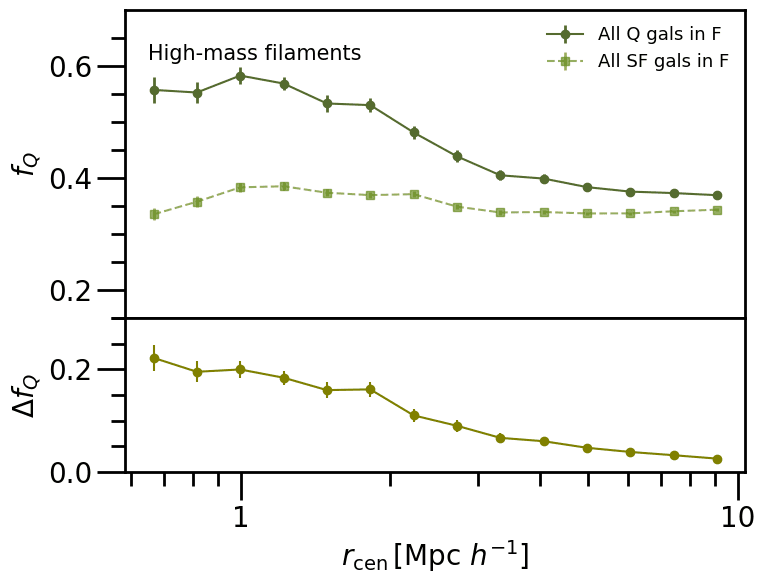}
  \end{subfigure}
  \caption{Quenched fraction $f_{\rm Q}$ of neighboring galaxies around all low-mass central galaxies in low-mass (left) and high-mass (right) filamentary regions in TNG300. The lower panel shows the conformity signal, $\Delta f_{\rm Q}$, per each case.}
  \label{fig:conf_in_fil_per_mass_range}
\end{figure*}

Figure \ref{fig:conf_in_fil_per_mass_range} shows the quenched fraction around quenched and star-forming low-mass central galaxies that reside in low-mass (left panel) and high-mass (right panel) filaments, with the corresponding conformity signal, $\Delta f_{\rm Q}$, shown in the lower panels.

\FloatBarrier

We find that low-mass central galaxies residing in more massive filaments exhibit a systematically stronger conformity signal than their counterparts inhabiting lower-mass filaments. This result indicates that the amplitude of the conformity signal in filaments is sensitive to the local density traced by the filament mass.
While filamentary environments already play a role in shaping the residual conformity signal, the mass of the filament further modulates its strength. This behavior is consistent with an environmental origin of large-scale conformity, in which denser filamentary structures enhance the effectiveness of the external processes acting on low-mass central galaxies.

\section{Assessing the impact of merger candidate neighbors on the conformity signal} \label{sec:appendixG}

The distribution of M$_{200}$ for galaxies in different environments (see Fig. \ref{fig:m200_in_diff_environments}) reveals that quenched galaxies in SAG inhabit less massive DM halos than their star-forming counterparts, a trend that is not observed for quenched and star-forming galaxies in TNG300. Since former satellite galaxies have been removed from the samples and the analysis is performed separately by environment, these differences suggest that additional mechanisms may be acting on these galaxies.

Given that the SAG model contains a larger number of massive structures than TNG300, a plausible explanation for this behavior is the presence of tidal interactions between massive systems and nearby low-mass central galaxies. To test the role of these interactions and assess their potential impact on the conformity signal, we identify merger candidate systems in the semi-analytic model. These are defined as group-group, cluster-cluster, or group-cluster pairs that satisfy the condition $d \leq r_{1} + r_{2}$, where $d$ is the separation between the two massive systems, and $r_{1}$ and $r_{2}$ are the virial radii of each interacting system.  

\begin{figure*}[htbp]
  \centering
    \includegraphics[width=0.85\textwidth]{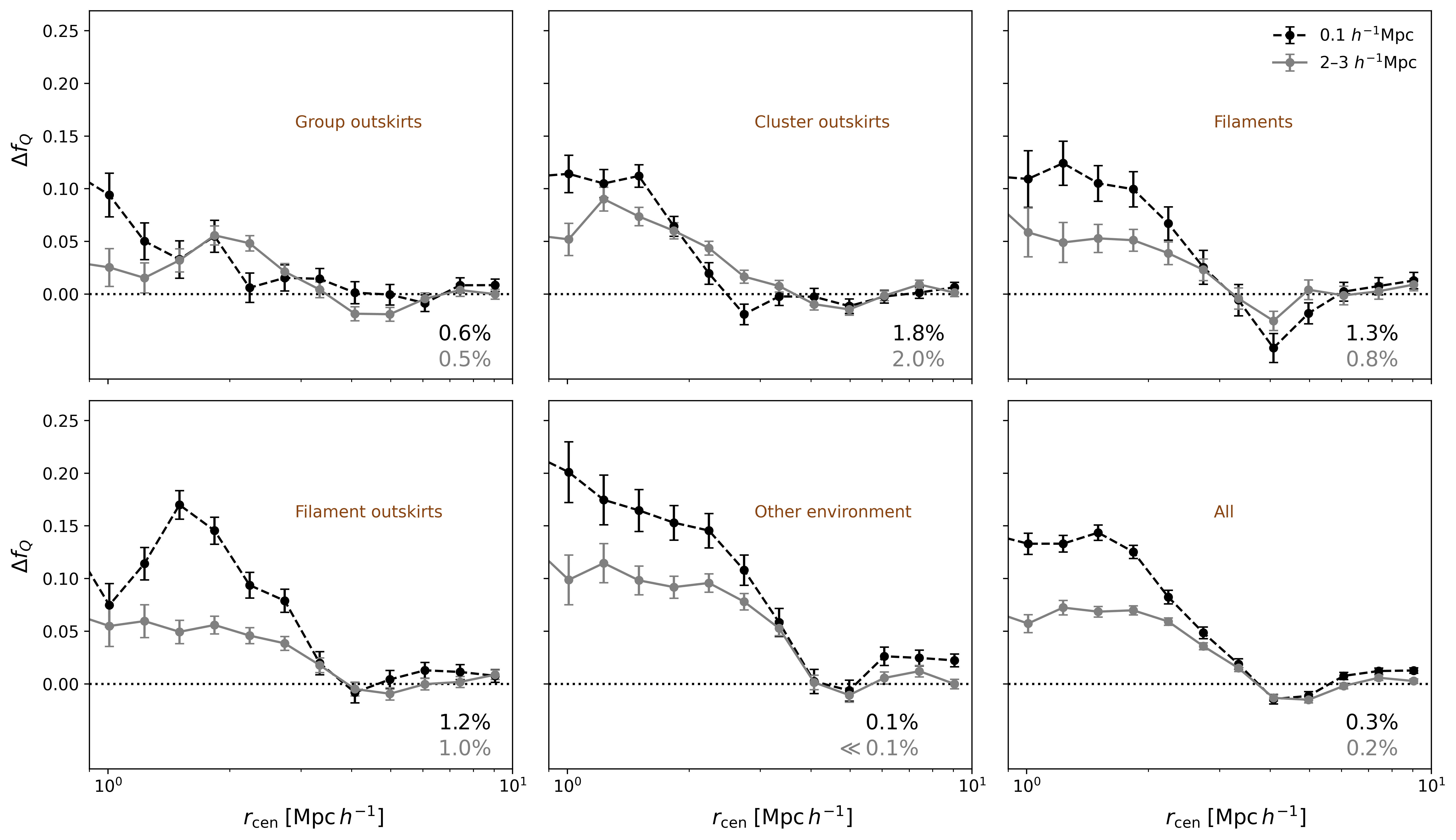}
  \caption{Conformity signal for low-mass central galaxies in different environments in SAG. The black dashed line shows the signal for galaxies with a nearby merger candidate system located at distances $\leq 0.1 \mathrm{Mpc}~h^{-1}$. The gray solid lines correspond to galaxies with a merger candidate system located between $2$ and $3 \mathrm{Mpc}~h^{-1}$. The percentages indicate the fraction of galaxies in each environment satisfying the merger candidate selection criteria.}
  \label{fig:conf_with_CM}
\end{figure*}

We search for merger candidate systems in two distinct distance ranges around low-mass central galaxies: a very close region at distances $\lesssim 0.1~h^{-1}$Mpc (to amplify as much as possible the effect), and an intermediate region between $2$ and $3~h^{-1}$Mpc. Figure \ref{fig:conf_with_CM} shows the conformity signal measured across different environments in SAG, for galaxies with a nearby merger candidate system at $\lesssim0.1~h^{-1}$Mpc (black dashed line), and for galaxies with a merger candidate system located at $2-3~h^{-1}$Mpc (gray solid lines). The percentages reported in each panel indicate the fraction of galaxies in each environment that satisfy the merger candidate selection criteria.

\FloatBarrier

We find a clear enhancement in the amplitude of the conformity signal across all environments, on average up to $4~h^{-1}$Mpc in both cases. The signal is stronger at small separations for galaxies with a nearby merger candidate system (black dashed line) than for those with a more distant interacting system (gray solid line). The signal amplification is more pronounced in filamentary regions and in other environments. Despite the small fraction of galaxies in each case, indicated by the percentages shown in each panel, we can identify how the signal is enhanced when central galaxies have an interacting massive neighbor. These results in addition exhibit that the influence of these systems on the signal reaches distances up to $4~h^{-1}$Mpc, independent of the distance of the interacting system with the low-mass central galaxy. The differences arises in the amplitude of the signal at lower separations.

We further evaluate this effect by considering the full galaxy sample, independently of environment, and recover the same overall trend. Galaxies located very close to merger-candidate systems ($<0.1~h^{-1}$Mpc) exhibit an enhanced conformity signal compared to those located at larger separations (2-3$~h^{-1}$Mpc) from these interacting massive systems. This result emphasizes the role of massive structure neighborhoods in shaping the large-scale conformity signal.

Although the conformity signal is enhanced in the vicinity of merger-candidate systems, we do not find significant differences in the median halo mass distributions between these two populations of galaxies. This indicates that the amplification of the signal is not driven by a halo mass bias but rather by dynamical environmental effects associated with tidal interactions.

\end{appendix}

\end{document}